\documentclass[aps,prc,twocolumn]{revtex4}

\newcommand{\snn}{\sqrt{s_{\rm _{NN}}}}
\newcommand{\tauf}{\tau_{\rm _F}}
\newcommand{\aT}{A_{\rm _T}}
\newcommand{\dETdy}{\frac{dE_{\rm _T}}{dy}}
\newcommand{\dETdytext}{dE_{\rm _T}/dy}
\newcommand{\yCM}{y_{\rm _{CM}}}
\newcommand{\dmTdy}{\frac{dm_{\rm _T}}{dy}}
\newcommand{\dmTdytext}{dm_{\rm _T}/dy}
\newcommand{\dNBdy}{\frac{dN_{\rm netB}}{dy}}
\newcommand{\dNBdytext}{dN_{\rm netB}/dy}
\newcommand{\mN}{m_{\rm _N}}
\newcommand{\nB}{n_{\rm _B}}
\newcommand{\nQ}{n_{\rm _Q}}
\newcommand{\nS}{n_{\rm _S}}
\newcommand{\muB}{\mu_{\rm _B}}
\newcommand{\muQ}{\mu_{\rm _Q}}
\newcommand{\muS}{\mu_{\rm _S}}
\newcommand{\tQGP}{t_{\rm _{QGP}}}

\usepackage{amsmath}
\usepackage{graphicx}
\usepackage[colorlinks,linkcolor=blue,urlcolor=blue,citecolor=blue]{hyperref}

\begin{document}

\title{Semi-analytical calculation of the trajectory of relativistic nuclear
collisions in the QCD phase diagram}

\author{Todd Mendenhall}
\email{mendenhallt16@students.ecu.edu}
\affiliation{Department of Physics, East Carolina University, Greenville, NC
27858}
\author{Zi-Wei Lin}
\email{linz@ecu.edu}
\affiliation{Department of Physics, East Carolina University, Greenville, NC
27858}

\date{\today}

\begin{abstract}
We extend a semi-analytical model that includes the finite nuclear
thickness to calculate the energy density $\epsilon(t)$ and
conserved-charge densities including the net-baryon density 
$n_{\rm _B}(t)$ produced at mid-spacetime-rapidity in central Au+Au 
collisions. Assuming the formation of a quark-gluon plasma with an
ideal gas equation of state of either quantum or Boltzmann statistics
or with a lattice QCD-based equation of state, we extract the
temperature $T$ and chemical potentials $\mu_{\rm _B}$, $\mu_{\rm _Q}$ 
and $\mu_{\rm _S}$ as functions of time. This then allows us to
semi-analytically calculate the $T-\mu_{\rm _B}$ trajectory of
relativistic nuclear collisions in the QCD phase diagram, which should
benefit the studies of high density physics including the search for
the critical end point. This model is also useful for exploring the
trajectories in the more general 
$T-\mu_{\rm _B}-\mu_{\rm _Q}-\mu_{\rm _S}$ QCD phase space. 
\end{abstract}

\maketitle

\section{Introduction}

Experiments at the Relativistic Heavy Ion Collider (RHIC) at
BNL~\cite{Jacobs2005} and the Large Hadron Collider (LHC) at
CERN~\cite{Tomasik2004} have produced the Quark-Gluon Plasma (QGP)
with ultrarelativistic nuclear collisions. 
The QGP is an exotic phase of matter in which the temperature and density
are high enough to ``melt'' hadrons~\cite{Adams2005a}. Such a system
of unbound quarks and gluons can only exist for a short time before
the partons recombine into hadrons due to confinement. The fleeting
nature of the QGP makes studying its properties very difficult, but
the success in this endeavor would expand our understanding of the
earliest stage of the universe during which the QGP is believed to
have existed. Learning about the QGP properties would also enable
the testing of Quantum ChromoDynamics (QCD) as the fundamental
theory governing the strong interaction~\cite{Aoki2006}. Of
particular interest to the field is understanding the phase transition
from hadronic to partonic matter in the QCD phase diagram
\cite{Bellwied2015}. Lattice QCD results show that it is a smooth
crossover at zero baryon chemical potential $\muB$
\cite{Bazavov2019}, but calculations at finite $\muB$ are difficult
\cite{Gavai2003}.

The conjectured critical end point (CEP) of the first-order phase
transition line is of special interest~\cite{Stephanov2005}. The Beam
Energy Scan (BES) program at RHIC uses Au+Au collisions at a variety
of energies to search for the
CEP~\cite{Aggarwal2010,STARCollaboration2020,Adamczyk2014}.
The matter created in a given collision system, for example, at a
given collision energy and centrality (or impact parameter), follows a
unique average trajectory and freezes out at a unique point on average
in the QCD phase diagram. Here, ``average'' refers to averaging over
many events of the given collision system. The time evolutions of the
temperature $T$ and baryon chemical potential $\muB$ together
determine the system's history in the QCD phase diagram. For those
collisions where the trajectory is near the CEP, event-by-event
fluctuations in conserved quantities could point to the existence of
the CEP~\cite{Koch2005}. For example, event-by-event net-proton
cumulant ratios at low collision energies could prove useful for
locating the CEP~\cite{Luo2017a}.

Since the matter in ultrarelativistic nuclear collisions progresses
through several stages, it is not straightforward to correlate
experimental measurements with the QGP thermodynamic properties.
Dynamical models including hydrodynamic models and
transport
models~\cite{Zhang2007,Lin2014a,Shen:2020mgh,Wang:2021owa}
have been used to study the evolution of the thermodynamic properties
of the QCD matter created during high energy nuclear collisions. Additionally,
semi-analytical models of the initial energy density
production~\cite{Bjorken1983,Lin2018,Mendenhall2021} have
progressively expanded our understanding of the early time evolution
of the energy densities produced in such collisions. In particular, it has
been shown that the effect of the finite nuclear thickness drastically alters
the peak energy density $\epsilon^{\rm max}$ at low collision energies
such as the BES energies~\cite{Lin2018,Mendenhall2021}.
  
The purpose of this study is to calculate trajectories in the QCD phase
diagram of the matter produced in central Au+Au collisions for collision
energies $\snn$ up to 200 GeV. Note that all of our results here
are for the central spacetime rapidity region ($\eta_s=0$), where our
semi-analytical model~\cite{Mendenhall2021} has been defined.
The paper is organized into the following sections after the
Introduction. First, we describe our semi-analytical
model~\cite{Mendenhall2021} for calculating the energy density
$\epsilon(t)$ and conserved-charge densities in
Sec. \ref{methods_1}. Second, we discuss the thermodynamic equations
governing an ideal gas of massless gluons and quarks under
quantum (Bose-Einstein and Fermi-Dirac) statistics in
Sec. \ref{methods_2} and Boltzmann statistics in
Sec. \ref{methods_3}. For the remainder of  this paper, we will refer
to the ideal gas equation of state (EoS) with quantum statistics as
the quantum EoS and the ideal gas EoS with Boltzmann statistics
as the Boltzmann EoS. For completeness, the
general relations between $\epsilon, n$ and $T, \mu$ that lead to the
thermodynamic equations of our semi-analytical model are provided in
the Appendix for both the quantum EoS and the Boltzmann EoS. Third, 
we introduce  in Sec. \ref{methods_4} a lattice QCD-based EoS, which
we will refer to as the lattice EoS, that provides a more realistic
relationship between $\epsilon, n$ and $T, \mu$ (at least at small $\muB$).  
We then present our results for $\epsilon(t)$
and $\nB(t)$ in Sec. \ref{results_1}, and give our results for
the extracted $T(t)$, $\muB(t)$, $\muQ(t)$, $\muS(t)$ and the resultant
$T-\muB$ trajectories for the quantum EoS in
Sec. \ref{results_2} and the Boltzmann EoS in
Sec. \ref{results_3}. In Sec. \ref{results_4}, we show the results for the
extracted trajectories by applying the lattice EoS to 
the densities calculated from our semi-analytical model. 
We then consider the effect of transverse
expansion on the trajectories in Sec. \ref{results_5}.
Results for the QGP lifetime for all three
equations of state are presented in Sec. \ref{results_6}. We then
discuss the improved net-baryon rapidity density parameterization, the 
effect of a finite $s$-quark mass, the near-zero value of the
calculated $\muQ$, and the implications of strangeness neutrality on
the results using the lattice EoS in Sec. \ref{discussions}. 
Finally, we conclude in Sec. \ref{conclusions}.

\section{Methods}
\label{methods}

\subsection{Calculating energy and net conserved-charge densities}
\label{methods_1}

Naively, one can use the method of the Bjorken energy density
formula~\cite{Bjorken1983}, in which partons at mid-rapidity are
produced at $t=0$ and $z=0$, to predict the time evolution of the
initial energy density $\epsilon(t)$:
\begin{equation}
\epsilon^{_{Bj}}(t)=\frac{1}{\aT~t}\dETdy.
\label{ep_bj}
\end{equation}
Here, $\aT=\pi R_A^2$ is the full transverse overlap area of two nuclei in
central A+A collisions, and $dE_{\rm T}/dy$ is the transverse energy
per rapidity at mid-rapidity. Because Eq.\eqref{ep_bj} predicts a diverging 
$\epsilon$ as $t\to 0$, one must choose a finite initial time
$\tauf$, which can be considered as the time when partons are
formed. For high collision energies, such as the top RHIC energy of
$\snn=200$ GeV, the finite thickness of the Lorentz-contracted nucleus
is small compared to the typical $\tauf$ value, so Bjorken's formula
is valid. However, for lower collision energies where the crossing time
$d_t=2R_A/\sinh{\yCM}$ is comparable to or even greater than
$\tauf$,  Eq.\eqref{ep_bj} is expected to break
down~\cite{Adcox2005}. Note that $\yCM$ is the rapidity of the 
projectile nucleus in the center-of-mass frame, and we use
$R_A=1.12A^{1/3}$ fm for the hard-sphere nuclear radius. 
Also note that the transverse expansion of the overlap volume is
neglected until Sec. \ref{results_5}. Furthermore, the 
slowing down of participant nucleons, and secondary parton or hadron
scatterings are neglected in our semi-analytical study, as done in
previous similar studies~\cite{Bjorken1983,Lin2018,Mendenhall2021}.

We have shown earlier~\cite{Lin2018,Mendenhall2021} that the finite
nuclear thickness must be considered when estimating $\epsilon(t)$ at
low collision energies. Therefore, a more realistic model of the initial
energy production is one in which partons are produced within
a finite range of time and longitudinal
position~\cite{Mendenhall2021}. This improved model has been shown to 
predict a finite $\epsilon^{\rm max}$ for $\tauf=0$ fm/$c$, while earlier 
models~\cite{Lin2018,Bjorken1983} predict infinite $\epsilon^{\rm max}$
there. In this model~\cite{Mendenhall2021}, the initial energy density
at time $t$ averaged over the full transverse overlap area is given by
\begin{equation}
\epsilon(t)=\frac{1}{\aT}
\iint_S\frac{dx\,dz_0}{t-x}\frac{d^3m_{\rm
    _T}}{dx\,dz_0\,dy}\cosh^3{\!y}.
\label{ep}
\end{equation}
In the above, $S$ represents the production area in the initial
production time $x$ and longitudinal position $z_0$ at observation
time $t$. The production area $S$ is the the portion of the overlap
region $S_0$ below the formation time hyperbola of Eq.~\eqref{zF}.
This ensures that a parton will contribute to the energy or
net-charge density at time $t$ after its formation time has
passed. The overlap region $S_0$ is a diamond along the vertical ($t$)
axis with one vertex at $(0,0)$ and the other at $(0,d_t)$ in the
$t-z$ plane. The width of the overlap region corresponds to the
thickness in the $z$-direction of the overlapping Lorentz-contracted
spherical nuclei. Therefore, the width $\Delta z$ at $t=0$ and $t=d_t$
are both zero, since these are the times when the nuclei would just
touch or completely pass through each other.
At $t=d_t/2$, the width is maximal at $\Delta z = 2R_A/\gamma$.
One key difference between our picture and that of the Bjorken energy
density formula is that the overlapping and expanding stages cannot be
clearly separated but are instead mixed together. 
We assume that partons free-stream from their production point
($z_0,x$) during time $t \in (x, x+\tauf \cosh{y})$ and could then interact
with the medium after its formation time. 
In our model, the velocity along the $z$-direction of a produced
parton is $v_z=(z-z_0)/(t-x)$, while in the Bjorken picture $v_z=z/t$
because all partons are produced at $(z_0,x)=(0,0)$. On the other
hand, our model is similar to the Bjorken energy density formula in
that all secondary interactions are ignored.

To be general, the initial energy is assumed to be produced
within  $x\in[t_1,t_2]$ (instead of the naive range $[0,d_t]$).
The resulting $\epsilon(t)$ is a piecewise function of $t$, where the
pieces are determined by the integration limits in Eq.\eqref{ep} and
are listed in Table \ref{tab_1} (see Ref.~\cite{Mendenhall2021} for
details). Note that $\epsilon(t)=0$ for $t\in[0,t_1+\tauf)$.

\begin{table*}
  \begin{tabular}{cccc}
  \hline
  \hline
  Piece & $t$-range & $x$-range & $z_0$-range \\
  \hline
	\\
  $\epsilon_{\rm _I}(t)$ or $n_{\rm _{B,I}}(t)$ & [$t_1+\tauf$, $t_a$)
	& [$t_1$, $x_1$) & [$-\beta(x-t_1)$, $\beta(x-t_1)$] \\
  & & [$x_1$, $t-\tauf$] & [$-z_{\rm _F}(x)$, $z_{\rm _F}(x)$] \\
	\\
  $\epsilon_{\rm _{II}}(t)$ or $n_{\rm _{B,II}}(t)$ & [$t_a$, $t_2+\tauf$)
	& [$t_1$, $t_{\rm mid}$) & [$-\beta(x-t_1)$, $\beta(x-t_1)$] \\
  & & [$t_{\rm mid}$, $x_2$) & [$-\beta(t_2-x)$, $\beta(t_2-x)$] \\
  & & [$x_2$, $t-\tauf$] & [$-z_{\rm _F}(x)$, $z_{\rm _F}(x)$] \\
	\\
  $\epsilon_{\rm _{III}}(t)$ or $n_{\rm _{B,III}}(t)$ & [$t_2+\tauf$, $\infty$)
	& [$t_1$, $t_{\rm mid}$) & [$-\beta(x-t_1)$, $\beta(x-t_1)$] \\
  & & [$t_{\rm mid}$, $x_2$] & [$-\beta(t_2-x)$, $\beta(t_2-x)$] \\
	\\
  \hline
  \hline
  \end{tabular}
\caption{Piecewise solution of $\epsilon(t)$ and $\nB(t)$ as functions
  of the observation time $t$, where the integration limits for each
  piece are written in the format $x\in[x^{\rm min},x^{\rm max}]$ and
  $z_0\in[z_0^{\rm min},z_0^{\rm max}]$ for each part of the production area
  in the initial production time $x$ and longitudinal position
  $z_0$~\cite{Mendenhall2021}.} 
  \label{tab_1}
\end{table*}

In this study, we assume that $d^3m_{\rm _T}/(dx\,dz_0\,dy)$ in
Eq.\eqref{ep} is factorized:
\begin{equation}
\frac{d^3m_{\rm _T}}{dx\,dz_0\,dy}=g(z_0,x)\dmTdy.
\label{d3mT}
\end{equation}
The weighting function $g(z_0,x)$ is normalized as
$\iint_{S_0}g(z_0,x)dx\,dz_0=1$, where $S_0$ is the area of the entire
diamond-shaped production region in the $t-z$
plane~\cite{Mendenhall2021}. This normalization condition ensures that
$\dmTdytext$ represents the initial transverse mass rapidity density
of all produced partons. We further make the simplest assumption that
partons are produced uniformly throughout $S_0$, which leads to 
$g(z_0,x)=2/(\beta t_{21}^2)$, where $\beta=\tanh{\yCM}$ and
$t_{21}=t_2-t_1$. Next, we take the following
specific form for $\dmTdytext$~\cite{Mendenhall2021}:
\begin{equation}
\dmTdy=\dETdy+\mN\dNBdy, 
\label{dmtdy}
\end{equation}
where $\mN$ is the nucleon mass. 
We assume that the $\dETdytext$ term is described by a single Gaussian
function while $\dNBdytext$ is described by a double
Gaussian~\cite{Mendenhall2021}: 
\begin{align}
\dETdy&=\dETdy(0)~
\exp\!{\left(\!-\frac{y^2}{2\sigma_1^2}\!\right)},
\label{dETdy}\\
\dNBdy&\propto
\exp\!{\left[-\frac{(y+y_{\rm _B})^2}{2\sigma_2^2}\right]}
+\exp\!{\left[-\frac{(y-y_{\rm _B})^2}{2\sigma_2^2}\right]}.
\label{dNBdy}
\end{align}
The transverse energy rapidity density at mid-rapidity is
parameterized as $dE_{\rm _T}/dy(0)=1.25\,dE_{\rm _T}/d\eta(0)$, where
$dE_{\rm _T}/d\eta(0)$ at $\snn>20.7$ GeV is taken as the
parameterization given by the PHENIX Collaboration~\cite{Adler}, while
$dE_{\rm  _T}/d\eta(0)$ at lower energies is given by an improved 
parameterization~\cite{Mendenhall2021}. 
We use the following parameterizations for the Gaussian
parameters $y_{\rm _B}$ and $\sigma_2$:
\begin{align}
\begin{split}
&y_{\rm _B}=0.599\left[1-\frac{1}{2.18+\ln^{1.86}\left(\frac{\snn}{E_0}\right)}\right]\yCM, \\
&\sigma_2=0.838\left[1-\frac{1}{5.01+\ln^{1.61} \left(
      \frac{\snn}{E_0}\right)}\right] \sqrt{\ln\left(\frac{\snn}{E_0}\right)}
\end{split}
\label{ybs2}
\end{align}
with $E_0=2\mN$ being the threshold energy. 
Note that these parameterizations are different from those used in our
earlier study~\cite{Mendenhall2021}, and we explain our reasoning in 
Sec.~\ref{discussions}. 
These parameterizations have been obtained using the proton $dN/dy$
data at $\snn=$ 2.65, 3.30, 3.85, and 4.31 GeV~\cite{Klay2002} and the
net-proton $dN/dy$ data at $\snn=$
4.87~\cite{E802:1999hit,E917:2000spt}, 8.77~\cite{NA49:2010lhg},
17.3~\cite{NA49Collaboration1998}, 62.4~\cite{BRAHMS:2009wlg} and 200
GeV~\cite{BRAHMSCollaboration2003} in central Au+Au collisions 
(with the exception that central Pb+Pb data are used at 8.8 and 17.3 GeV).
The value of the proportionality constant in Eq.\eqref{dNBdy} is
determined from the conservation of total net-baryon number
$\int(\dNBdytext)\,dy=2A$ at each collision energy. Finally, the
Gaussian parameter $\sigma_1$ in Eq.\eqref{dETdy} is calculated using
the conservation of total energy
$\int(\dmTdytext)\cosh{\!y}\,dy=A\snn$ at each collision energy. 

In our semi-analytical model~\cite{Lin2018,Mendenhall2021}, the
primary collisions between the two nuclei start at time $t_1$ and end
at time $t_2$. In this study, we take
\begin{equation}
t_1=\frac{1}{6}d_t,~t_2=\frac{5}{6}d_t,
\label{t1t2}
\end{equation}
because this choice gives $\epsilon^{\rm max}=2\rho_0\mN$ and
$n_{\rm B}^{\rm max}=2\rho_0$ for the threshold collision energy $\snn=E_0$,
which would be expected if the two nuclei would just fully overlap.
Note that $\rho_0\approx0.17$ fm$^{-3}$
in the hard sphere model for the nucleus.
Also note that in previous studies~\cite{Lin2018,Mendenhall2021},
$t_1=0.2d_t$ and $t_2=0.8d_t$ were used so that the width of the
production time distribution was similar to the results from the
string melting version of a multi-phase transport (AMPT)
model~\cite{Lin2005}. In Table~\ref{tab_1}, time $t_a$ is given by 
\begin{equation}
t_a=t_{\rm mid}+\sqrt{\tauf^2+\left(\!\frac{\beta t_{21}}{2}\!\right)^{\!2}}
\label{ta}
\end{equation}
with $t_{\rm mid}=(t_1+t_2)/2$, times $x_1$ and $x_2$ are given by
\begin{equation}
x_i=\frac{t-\beta^2t_i-\sqrt{\beta^2[(t-t_i)^2-\tauf^2]+\tauf^2}}{1-\beta^2}
\label{xi}
\end{equation}
for $i=1$ or 2, and the function $z_{\rm F}(x)$ is given by
\begin{equation}
z_{\rm F}(x)=\sqrt{(t-x)^2-\tauf^2}.
\label{zF}
\end{equation}
  
We now calculate the net-baryon density $\nB(t)$ using the same method
as that used for the $\epsilon(t)$
calculation~\cite{Mendenhall2021}. We then obtain the following
equation for the net-baryon density that is similar to Eq.\eqref{ep}:
\begin{equation}
\nB(t)=\frac{1}{\aT}\iint_S\frac{dx\,dz_0}{t-x}
\frac{d^3N_{\rm netB}}{dx\,dz_0\,dy}\cosh^2{\!y}.
\label{nB_dens}
\end{equation}
Note that there is one less power of $\cosh{\!y}$ in this equation than
in Eq.\eqref{ep} because that equation involves $E=m_{\rm _T}\cosh{\!y}$.
We also assume the same factorization $d^3N_{\rm
  netB}/(dx\,dz_0\,dy)=2/(\beta t_{21}^2)\dNBdytext$. 
Therefore, the net-baryon density $\nB(t)$ is also given by a
piecewise solution, as shown in Table~\ref{tab_1}.

Since the net-electric charge is carried by the incoming protons while
the net-baryon number is carried by the incoming nucleons in the
nuclei, we assume that the initial production from the primary NN
collisions is independent of whether N is a proton or a neutron.
Our semi-analytical method then leads to
\begin{equation}
	\nQ(t)=\nB(t)\frac{Z}{A},
\end{equation}
where $Z$ and $A$ represent the atomic number and mass number of the
nucleus, respectively, in the symmetric A+A system. Note that
the relationship $\nQ/\nB=Z/A$ has also been used in other
studies~\cite{Monnai:2021kgu}.
Furthermore, since the incoming nuclei do not carry net-strangeness,
we assume that the initial production is symmetric for $s$ and $\bar
s$. 
For the net-strangeness density, our semi-analytical method then simply gives
\begin{equation}
\nS(t)=0, 
\end{equation}
i.e., strangeness neutrality. 

\subsection{Thermodynamics of a massless QGP with the quantum EoS} 
\label{methods_2}

According to Eq.\eqref{nS_q} or Eq.\eqref{nS_b}, the result $\nS=0$
from our semi-analytical model gives the following relation
\begin{equation}
\muB-\muQ-3\muS=0
\label{mu_relation}
\end{equation}
for the ideal gas equation of state with 
either quantum or Boltzmann statistics, which corresponds to
$\mu_s=0$ for the strange quark chemical potential. Using this
relation, the general results in Eqs.\eqref{ep_q}-\eqref{nQ_q} for
quantum statistics simplify to the following set of equations:
\begin{align}
\begin{split}
&\epsilon=\frac{19\pi^2}{12}T^4+3\frac{(\muB-2\muS)^2+\muS^2}{2}T^2 \\
&\quad+3\frac{(\muB-2\muS)^4+\muS^4}{4\pi^2},
\end{split}\label{full_q_e} \\
&\nB=\frac{\muB-\muS}{3}T^2+\frac{(\muB-2\muS)^3+\muS^3}{3\pi^2},
\label{full_q_nB} \\
&\nQ=\frac{2\muB-5\muS}{3}T^2+\frac{2(\muB-2\muS)^3-\muS^3}{3\pi^2}.
\label{full_q_nQ}
\end{align}
We refer to the $T$, $\muB$, $\muQ$ and $\muS$ values extracted
from the $\epsilon$, $\nB$, and $\nQ$ values using
Eqs.\eqref{mu_relation}-\eqref{full_q_nQ} as the ``full solution''
for the quantum ideal gas EoS.

On the other hand, if one ignores the electric charge by setting
$\muQ=0$, Eq.\eqref{mu_relation} gives $\muS=\muB/3$, which leads
to the following simplified equations:
\begin{align}
  &\epsilon_{_1}=\frac{19\pi^2}{12}T^4+\frac{\muB^2}{3}T^2+\frac{\muB^4}{54\pi^2},
  \label{partial_1_q_e} \\
  &n_{\rm _{B,1}}=\frac{2\muB}{9}T^2+\frac{2\muB^3}{81\pi^2}.
  \label{partial_1_q_nB}
\end{align}
We refer to the $T$ and $\mu$ values extracted
from the $\epsilon$ and $\nB$ values using
Eqs.\eqref{partial_1_q_e}-\eqref{partial_1_q_nB} as the ``partial-1
solution'' for the quantum EoS.
However, note that in this case Eq.\eqref{full_q_nQ} would give
$\nQ=\nB/2$, which is inconsistent with the result $\nQ=\nB Z/A$
from our semi-analytical model. This discrepancy, which also exists
for the Boltzmann statistics, is a consequence of the choice for $\muQ$. On the
other hand, for the Au+Au collisions that we consider in this study,
$Z/A\approx0.4$ is not far from 1/2.

Additionally, if one ignores both the electric charge and strangeness
by setting $\muQ=\muS=0$, Eqs.\eqref{ep_q}-\eqref{nB_q} lead to
the following different set of simplified equations:
\begin{align}
  &\epsilon_{_2}=\frac{19\pi^2}{12}T^4+\frac{\muB^2}{2}T^2+\frac{\muB^4}{36\pi^2},
  \label{partial_2_q_e} \\
  &n_{\rm _{B,2}}=\frac{\muB}{3}T^2+\frac{\muB^3}{27\pi^2}.
  \label{partial_2_q_nB}
\end{align}
We refer to the $T$ and $\mu$ values extracted from the $\epsilon$ and
$\nB$ values using Eqs.\eqref{partial_2_q_e}-\eqref{partial_2_q_nB} as
the ``partial-2 solution'' for the quantum EoS.
Note that this approximation is inconsistent
with Eq.\eqref{mu_relation} from our semi-analytical model.
In addition, in this case Eqs.\eqref{nQ_q}-\eqref{nS_q} would give
$\nQ=0$ and $\nS=-\nB$ (same for the Boltzmann statistics), which are
more inconsistent with the results from our semi-analytical
model~\cite{Mendenhall2021} and the numerical results from the AMPT
model study~\cite{Wang:2021owa}.

\subsection{Thermodynamics of a massless QGP with the Boltzmann EoS}
\label{methods_3}

Using Eq.\eqref{mu_relation} that is also valid for Boltzmann
statistics, the general results in Eqs.\eqref{ep_b}-\eqref{nS_b} simplify
to the following set of equations:
\begin{align}
  \begin{split}
    &\epsilon=\frac{12}{\pi^2}T^4\Bigg[7+3\cosh{\left(\frac{\muB-2\muS}{T}\right)} \\
    &\qquad+3\cosh{\left(\frac{\muS}{T}\right)}\Bigg],
  \end{split} \label{full_b_e} \\
  &\nB=\frac{4}{\pi^2}T^3\Bigg[
  \sinh{\left(\frac{\muB-2\muS}{T}\right)}
  +\sinh{\left(\frac{\muS}{T}\right)}\Bigg], \\
  &\nQ=\frac{4}{\pi^2}T^3\Bigg[
  2\sinh{\left(\frac{\muB-2\muS}{T}\right)}
  -\sinh{\left(\frac{\muS}{T}\right)}\Bigg].
  \label{full_b_nQ}
\end{align}
They provide the full solution of $T$ and $\mu$ for the Boltzmann ideal gas EoS.

Again, if we ignore electric charge by setting $\muQ=0$, we then have
the following that defines the partial-1 solution for the Boltzmann EoS:
\begin{align}
  &\epsilon_{_1}=\frac{12}{\pi^2}\,T^4
	\left[7+6\cosh{\left(\frac{\muB}{3T}\right)}\right],
  \label{partial_1_b_e} \\
  &n_{\rm _{B,1}}=\frac{8}{\pi^2}\,T^3\sinh{\left(\frac{\muB}{3T}\right)}.
  \label{partial_1_b_nB}
\end{align}
Alternatively, if one sets $\muQ=\muS=0$,
one would simplify Eqs.\eqref{ep_b}-\eqref{nB_b}
to the following equations that determine the partial-2 solution for
the Boltzmann EoS:
\begin{align}
  &\epsilon_{_2}=\frac{12}{\pi^2}\,T^4
	\left[4+9\cosh{\left(\frac{\muB}{3T}\right)}\right],
  \label{partial_2_b_e} \\
  &n_{\rm _{B,2}}=\frac{12}{\pi^2}\,T^3\sinh{\left(\frac{\muB}{3T}\right)}.
  \label{partial_2_b_nB}
\end{align}

\subsection{Thermodynamics of the Lattice EoS}
\label{methods_4}

A lattice QCD-based
EoS~\cite{Borsanyi:2018grb,Noronha-Hostler:2019ayj} provides another 
way to relate the energy density and net conserved-charge densities
$\epsilon, n$ to the temperature and conserved-charge chemical
potentials $T, \mu$. First, the pressure is written as a Taylor series
in all three $\mu/T$ up to a total power 
$i+j+k\leq 4$~\cite{Noronha-Hostler:2019ayj}:
\begin{equation}
\frac{p}{T^4}=\sum_{i,\,j,\,k}\frac{1}{i!\,j!\,k!}\chi_{ijk}^{BQS}
\left(\!\frac{\muB}{T}\!\right)^{\! i} 
\left(\!\frac{\muQ}{T}\!\right)^{\! j}\left(\!\frac{\muS}{T}\!\right)^{\! k}.
\end{equation}
The coefficients $\chi_{ijk}^{BQS}$ have been calculated on a
$48^3\times12$ lattice in the temperature range $T\in[135,220]$
MeV~\cite{Borsanyi:2018grb}. However, this temperature range is not
enough to cover the full hydrodynamical evolution of the matter
produced in heavy-ion collisions, so a more complete EoS is
constructed~\cite{Noronha-Hostler:2019ayj}.  
The coefficients are smoothly merged with the results from the
hadron resonance gas model~\cite{Huovinen:2017ogf} to constrain the low
temperature behavior of the EoS, while in the high temperature regime
each coefficient is imposed to smoothly approach its Boltzmann
limit~\cite{Noronha-Hostler:2019ayj}.  Except for $\chi_{200}^{BQS}$,
each coefficient has been written as a ratio of ninth-degree
polynomials in inverse powers of a scaled temperature $T_1 \equiv
T/(154\,\rm {MeV})$:
\begin{equation}
\chi_{ijk}^{BQS}=
\frac{\sum_na^{ijk}_n/T_1^n}{\sum_nb^{ijk}_n/T_1^n}+c^{ijk}_0.
\label{chiijk}
\end{equation}
The susceptibility $\chi_{200}^{BQS}$ has a different form:
\begin{equation}
\chi_{200}^{BQS}=
e^{-h_1/T_2-h_2/T_2^2}f_3\left[1+\tanh{(f_4T_2+f_5)}\right],
\label{chi200}
\end{equation}
where $T_2 \equiv T/(200\,\rm {MeV})$~\cite{Noronha-Hostler:2019ayj}
is a different scaled temperature. In Eqs.\eqref{chiijk} and
\eqref{chi200}, $a_n$, $b_n$, $c_0$, $h_1$, $h_2$, $f_3$, $f_4$, and
$f_5$ are constant coefficients whose values are
published~\cite{Noronha-Hostler:2019ayj}.

With these parameterizations of $\chi_{ijk}^{BQS}$, the pressure is
fully defined, and one can use the standard thermodynamic relations to
find $\epsilon$, $\nB$, $\nQ$, and $\nS$, and the entropy density $s$ in
terms of $T$ and $\mu$:
\begin{equation}
\begin{split}
\frac{\epsilon}{T^4}&=
\frac{s}{T^3}-\frac{p}{T^4}+\frac{\muB}{T}\frac{\nB}{T^3}
+\frac{\muQ}{T}\frac{\nQ}{T^3}+\frac{\muS}{T}\frac{\nS}{T^3}, \\
\frac{\nB}{T^3}&=
\frac{1}{T^3}\frac{\partial p}{\partial \muB}\Bigg\rvert_{T,\muQ,\muS},~
\frac{\nQ}{T^3}=
\frac{1}{T^3}\frac{\partial p}{\partial \muQ}\Bigg\rvert_{T,\muB,\muS}, \\
\frac{\nS}{T^3}&=
\frac{1}{T^3}\frac{\partial p}{\partial \muS}\Bigg\rvert_{T,\muB,\muQ},~
\frac{s}{T^3}=
\frac{1}{T^3}\frac{\partial p}{\partial T}\Bigg\rvert_{\muB,\muQ,\muS}.
\end{split}
\label{lattice_EOS}
\end{equation}

When using this lattice EoS, we impose the conditions relevant to
heavy-ion collisions, which are already implemented in our
semi-analytical model~\cite{Lin2018,Mendenhall2021}: 
\begin{equation}
\nQ=\nB Z/A,\ \nS=0.
\label{condition}
\end{equation}
We can then use our $\epsilon(t)$ and $n(t)$ values as inputs to the
lattice EoS to extract the corresponding $T$ and $\mu$ values, which
we refer to as the full solution with the lattice EoS. 

Our method of extracting the $T$ and $\mu$ values for given
$\epsilon$ and $n$ values involves calculating the intersection of the
corresponding constant $\epsilon$ and $\nB$ contours in the $T-\muB$
plane. First we reject any solution that has $T<70$ MeV because in 
this region we find that the energy density reconstructed with 
Eq.\eqref{lattice_EOS} can be negative, which is unphysical.
Another complication is that there are often multiple solutions for
the $T$ and $\mu$ values (for a given set of $\epsilon$ and $n$ values).
There is often a branch of solutions at $\muB\gtrsim2$ GeV and
$T\gtrsim500$ MeV; we reject these solutions since they occur well
beyond the expected region of validity of the lattice QCD calculations 
($\muB/T\lesssim2.5$)~\cite{Noronha-Hostler:2019ayj}. 
At lower collision energies, the numerical solutions of the trajectory
often form two branches: the first behaving as expected, while the
second lies at lower $T$ and higher $\muB$. We reject the second
branch because it has larger $\muB/T$ values; in addition, that branch
occurs near or below $T\sim135$ MeV where the lattice QCD calculations
stop~\cite{Borsanyi:2018grb}. As a result, at $\snn=2.0$ GeV we find no 
$T$ and $\mu$ solutions for the densities at any time during the
evolution. For $\snn \gtrsim 4.0$ GeV, the trajectory can be extracted
with the lattice EoS, although at low energies (i.e., energies not
much higher than 4 GeV) only a portion of the trajectory around the
time of  $\epsilon^{\rm max}$ can be extracted.

\section{Results}
\label{results}

\subsection{Energy and net-baryon density}
\label{results_1}

\begin{figure*}
\centering
\includegraphics[width=\linewidth]{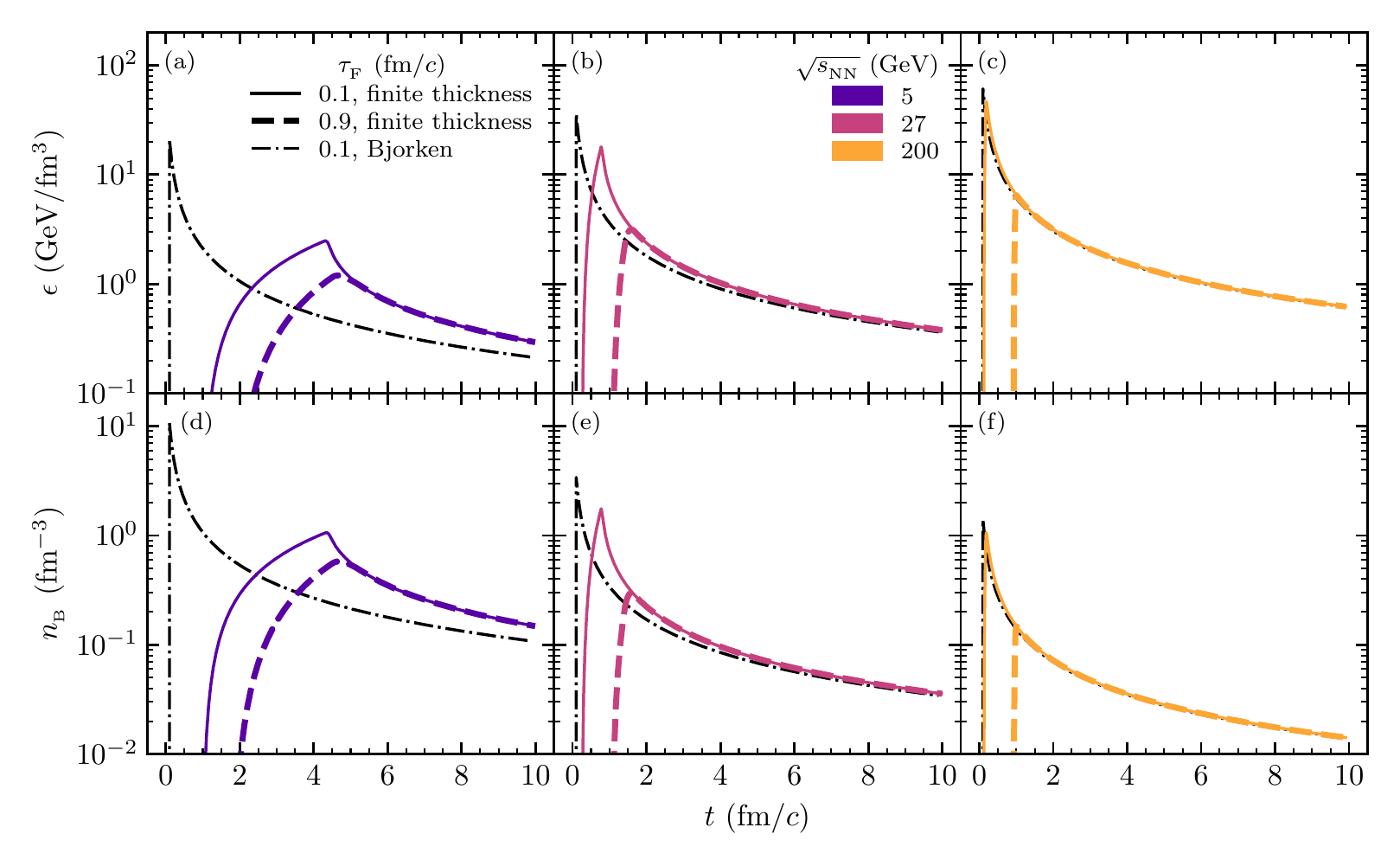}
\caption{(a-c) Energy density $\epsilon(t)$ and (d-f) net-baryon
  density $\nB(t)$ at mid-spacetime-rapidity averaged over the full
  transverse area for central Au+Au collisions at $\snn=$ 5.0, 27,
  and 200 GeV from our model with $\tauf=$ 0.1 (solid) and 0.9
  (dashed) fm/$c$ in comparison with results from the Bjorken formula 
  with $\tauf=$ 0.1 fm/$c$ (dot-dashed).} 
\label{fig_1}
\end{figure*}

Figure~\ref{fig_1} shows the results of $\epsilon(t)$ and $\nB(t)$
at mid-spacetime-rapidity averaged over the full transverse overlap
area for central Au+Au collisions at three different energies. 
The results from our semi-analytical model at $\tauf=$ 0.1 and 0.9
fm/$c$ are shown together with the results from the Bjorken energy
density formula at $\tauf=$ 0.1 fm/$c$. 
We observe that the peak energy density $\epsilon^{\rm max}$ from our model
increases as $\snn$ increases, while the peak net-baryon density
$\nB^{\rm max}$ mostly decreases with $\snn$ (except for 
an increase at low to moderate $\snn$ and small $\tauf$). 

We also see the large effect that the finite nuclear
thickness has on the predicted
densities at low collision energies~\cite{Lin2018,Mendenhall2021}.
Compared to the results from the Bjorken formula~\cite{Bjorken1983}, our
results from Eq.\eqref{ep} and Eq.\eqref{nB_dens} have significantly
smaller $\epsilon^{\rm max}$ and $\nB^{\rm max}$ at low collision energies,
while the difference from the Bjorken results decreases and
eventually vanishes at high collision energies as expected. Note that
we use $\dmTdytext$ of Eq.\eqref{dmtdy} for the $\dETdytext$ term in
Eq.\eqref{ep_bj} to calculate the Bjorken energy density as
$\epsilon^{_{Bj}}(t)=1/(\aT t) \; \dmTdytext$, and the
net-baryon density in the Bjorken formula is given by 
\begin{equation}
\nB^{_{Bj}}(t)=\frac{1}{\aT\,t}\dNBdy. 
\label{nB_bj}
\end{equation}
Figure~\ref{fig_1} also shows the $\tauf$-dependence of the densities, 
where $\epsilon^{\rm max}$ and $\nB^{\rm max}$ are lower and occur
later in time at a larger $\tauf$. 
In addition, the late time evolution of the densities approaches the
Bjorken results for all $\tauf$ and does so earlier in time for higher
collision energies. Therefore, we expect 
significant differences in the $T-\muB$ trajectories at low energies
between our results and the results from the Bjorken formula.
Note that at late times ($t \geq t_2 + \tauf$) our energy density and
net-baryon densities do not depend on the formation time, because the
integration limits of $x$ and $z_0$ at late times are independent of
$\tauf$ as shown in piece III of Table~\ref{tab_1}. 
Also, due to $v_z=(z-z_0)/(t-x)$, at late times only partons with
$y\sim 0$ will contribute to the densities at $z=0$ (regardless of
their production point), which is the same as for the
Bjorken densities. Therefore, at late times our densities approach the
results from the Bjorken density formulae.

We also note that the densities of our semi-analytical model can
decrease at a faster rate just after the time of maximum density than
the densities from the Bjorken formulae; this is more noticeable at
lower collision energies. The decrease of our energy and net-baryon
densities with time can be understood analytically because they
are similar to the results calculated using a uniform time
profile~\cite{Lin2018}. There, the densities decrease with time as
$\ln[(t-t_1)/(t-t_2)]$ for $t\geq t_2+\tauf$ after the maximum 
density is reached. Therefore, at the beginning of the decreasing part
of the curve (i.e., at $t$ just after $t_2+\tauf$), our densities
decrease as $\ln[1/(t-t_2)]$, which is faster than the $1/t$ behavior
of the Bjorken densities at the same $t$. On the other hand, this can
also be seen as a simple shift of time (i.e., from $t$ to $t-t_2$) due
to the finite nuclear thickness.

In addition, we notice that the maximum density at a given collision
energy for a larger $\tauf$ is almost on the tail part of the density
curve for a smaller $\tauf$. Again, we can understand this  
merging of the late time evolutions of the energy and net-baryon
densities by considering the case of the densities calculated using a
uniform time profile~\cite{Lin2018}.
At late times after the time of maximum density, the uniform time
profile gives densities which are proportional to
$\ln[(t-t_1)/(t-t_2)]$, which does not involve $\tauf$. Therefore, the
late time evolutions from our model are independent of $\tauf$. 

\subsection{Trajectory of a massless QGP with the quantum EoS}
\label{results_2}

\begin{figure*}
\centering
\includegraphics[width=\linewidth]{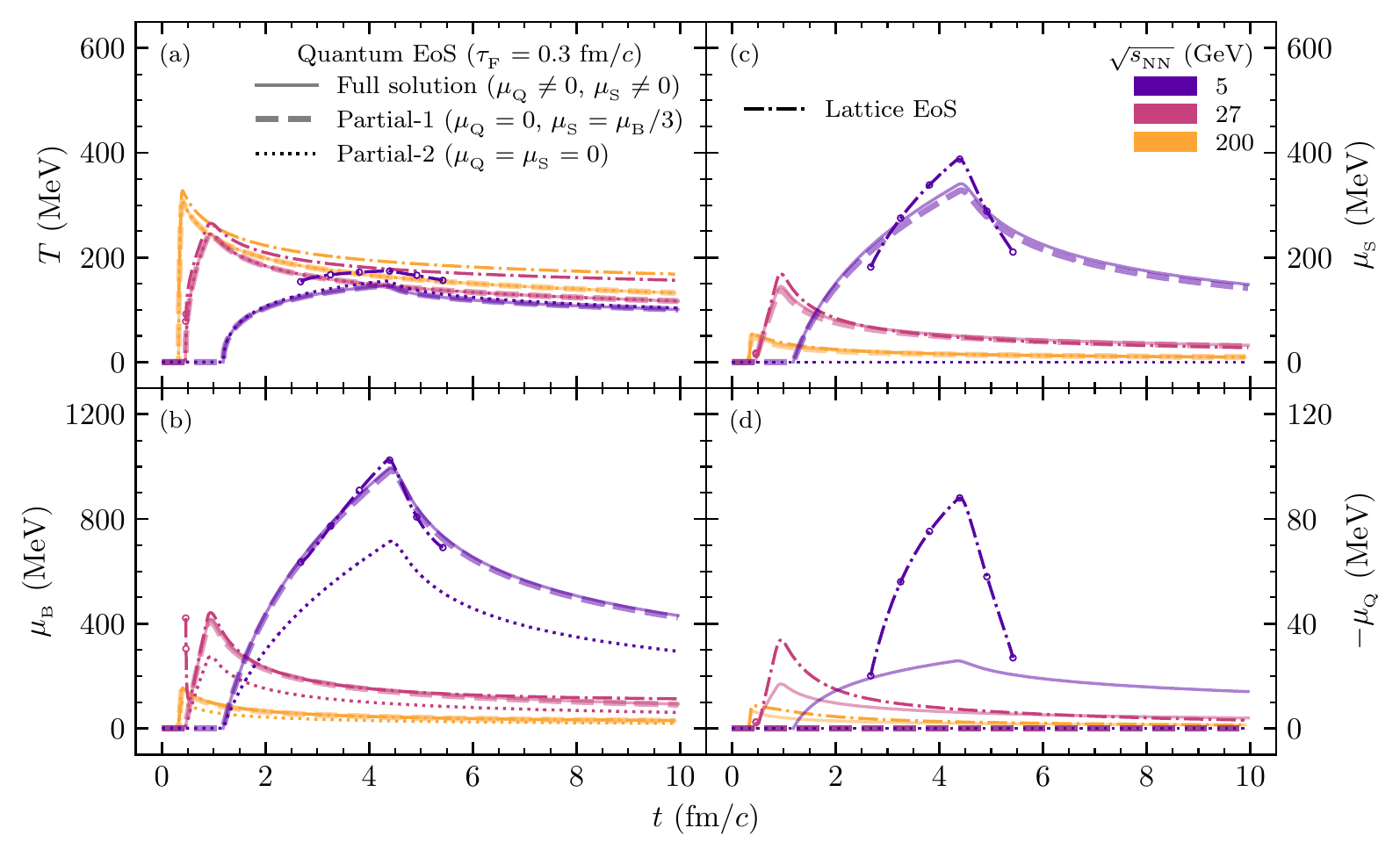}
\caption{(a) $T(t)$, (b) $\muB(t)$, (c) $\muS(t)$, and (d)
  $-\muQ(t)$ for a massless QGP with the quantum EoS using the full
  (solid), partial-1 (dashed), and partial-2 (dotted) solutions in
  comparison with results using the lattice EoS (dot-dashed). 
Open circles on the lattice curves represent times for which the
corresponding trajectory has $\muB/T>2.5$ and thus may be unreliable. 
All results are for central Au+Au collisions at $\snn=$ 5.0,
27, and 200 GeV with $\tauf=$ 0.3 fm/$c$.}
\label{fig_2}
\end{figure*}

In Fig.~\ref{fig_2}, we show the $T$ and $\mu$ results extracted using
the full solution, the partial-1 solution, and the partial-2 solution
of the quantum EoS, in comparison with the results extracted using 
the lattice EoS. 
We see that $T^{\rm max}$ increases with $\snn$, while $\muB^{\rm max}$
decreases with $\snn$ (from 5 GeV) for all four equations of state. It
is obvious that both $T$ and $\muB$ reach the peak value earlier
in time at higher energies due to the shorter duration time $d_t$. 
Figure~\ref{fig_2} also shows that the partial-1 solution
reproduces the $T(t)$ and $\muB(t)$ results from the full solution
almost exactly over the entire time evolution. On the other hand, the
partial-2 solution gives a much smaller $\muB$ than the full solution.  
In addition, we see that the magnitude of $\muQ$ in the full solution
is very small ($\lesssim 30$ MeV), so the assumption $\muQ=0$ in both
the partial-1 and partial-2 solutions is reasonable for the ideal gas
EoS. Furthermore, $\muS\simeq\muB/3$ in the full solution
results explains why the partial-1 solution that assumes
$\muS=\muB/3$ is much better than the partial-2 solution
that assumes $\muS=0$. We note that the recent numerical results 
from the AMPT model for the ideal gas EoS~\cite{Wang:2021owa} also show
$\muQ\approx0$ and $\muS\approx\muB/3$.

\begin{figure}
\centering
\includegraphics[width=\linewidth]{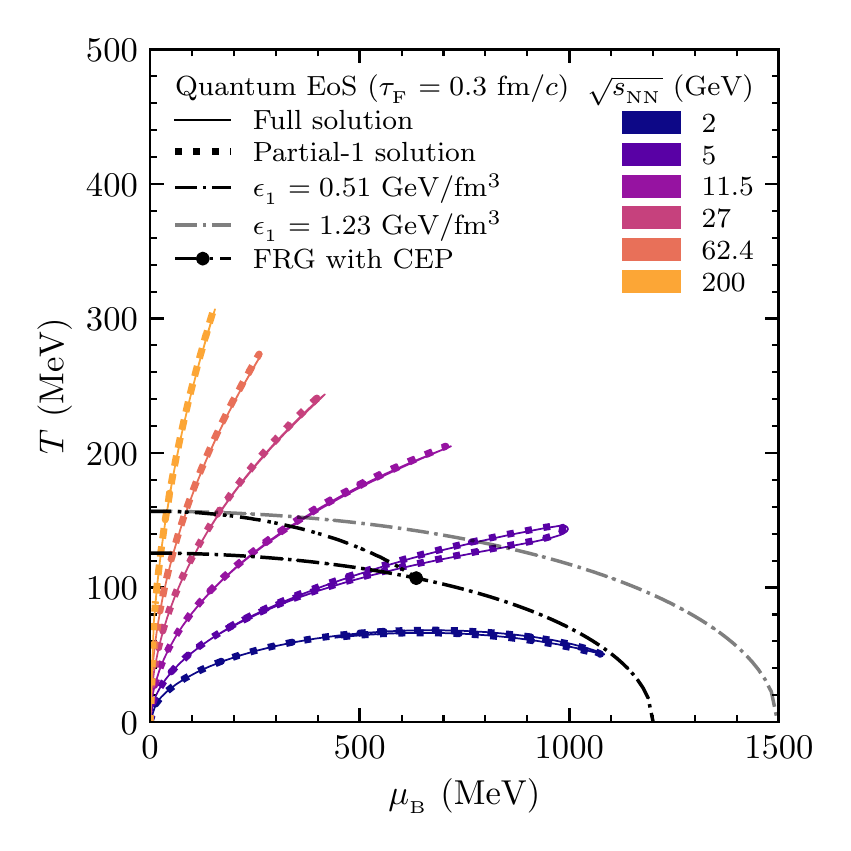}
\caption{Trajectories for a massless QGP with the quantum EoS for the
  full (solid) and partial-1 (dotted) solutions with $\tauf=$ 0.3
  fm/$c$ for central Au+Au collisions at $\snn=$ 2.0, 5.0, 11.5, 27.0,
  62.4, and 200 GeV. Two constant energy density lines from the
  partial-1   solution (dot-dashed) and the FRG crossover curve
  (dash-dot-dot) with the CEP (filled circle) are also shown for
  reference.}
\label{fig_3}
\end{figure}

Figure~\ref{fig_3} shows the $T-\muB$ trajectories in the QCD
phase diagram for the quantum EoS with $\tauf=$ 0.3 fm/$c$.
At very early times $t\in[0,t_1+\tauf)$, the system is at 
$(\epsilon,\nB)=(0,0)$, which corresponds to $(\muB,T)=(0,0)$ for the
ideal gas EoS. For $\snn\gtrsim4$ GeV, the trajectories pass through
the crossover curve, which comes from calculations using the
functional renormalization group (FRG) with $N_F=2+1$~\cite{Fu2020}. 
When a given trajectory reaches its endpoint, which corresponds to the time
when both $\epsilon^{\rm max}$ and $\nB^{\rm max}$ are reached, it turns
clockwise and returns toward the origin.
At high collision energies, the returning part of the trajectory
is so close to the outgoing part that the two appear to overlap.
At low energies such as $\snn=$ 5.0 GeV, however, the two
parts are distinguishable.
This behavior can be understood with the uniform time profile for the
density production~\cite{Lin2018}. 
In this case, both densities have the same time evolution and differ only
in their constants of proportionality.
Therefore, the time evolution of the
energy and net-baryon densities in our model are expected to be quite
similar at all times. 
Hence, the returning part of a trajectory should thus
overlap with the outgoing part because, 
for two points (one on the increasing part and the other on the
decreasing part) in a trajectory with the same energy density, 
the net-baryon densities will also be very similar. 

Notably, at $\snn\gtrsim 4.4$ GeV both the
outgoing and returning parts of the trajectory intersect the FRG
crossover curve, while at $\snn\lesssim 3.6$ GeV, neither part of the
trajectory intersects the FRG crossover curve. 
In Fig.~\ref{fig_3}, we
also see that the trajectories from the partial-1 solution match quite
closely those from the full solution. 
We have also used Eq.\eqref{partial_1_q_e} from the partial-1
solution for the quantum EoS to calculate the lines of constant
$\epsilon$ that go through the two endpoints of the FRG crossover
curve. They correspond to $\epsilon=$ 0.51 and 1.23 GeV/fm$^3$, as
shown in Fig.~\ref{fig_3}.
 
\begin{figure}
\centering
\includegraphics[width=\linewidth]{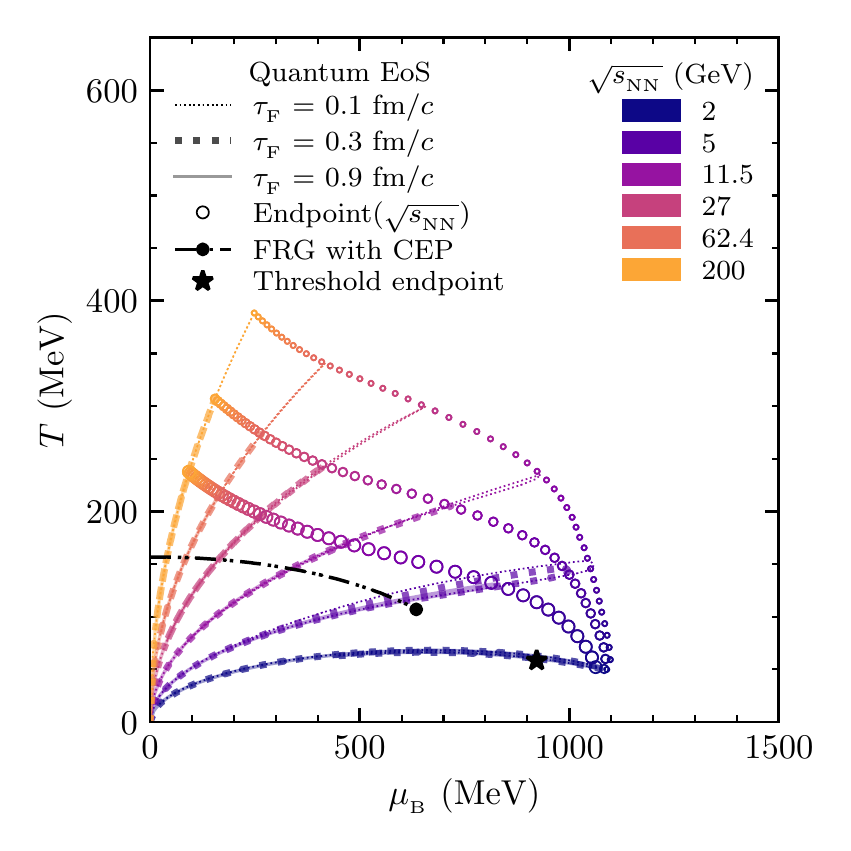}
\caption{Trajectories for the quantum EoS with $\tauf=$ 0.1 (thin
  dotted), 0.3 (thick dotted), and 0.9 (solid) fm/$c$ for central
  Au+Au collisions at $\snn=$ 2.0, 5.0, 11.5, 27.0, 62.4, and 200
  GeV, together with the trajectory endpoint curves (open circles) as 
  functions of $\snn$. The FRG crossover curve (dash-dot-dot) with the
  CEP (filled circle) and the endpoint at the threshold energy (star)
  are also shown for reference.} 
\label{fig_4}
\end{figure}

The results from our semi-analytical model~\cite{Mendenhall2021}
depend on the value of $\tauf$, and Fig.~\ref{fig_4} shows how the
$T-\muB$ trajectories from the quantum EoS full solution depend on
$\tauf$. As $\tauf$ decreases, the trajectory becomes longer with the
endpoint moving further to higher $\muB$ (and also higher $T$ except
at very low energies). Note that the trajectory endpoint corresponds
to both $\epsilon^{\rm max}$ and $\nB^{\rm max}$, but at low energies
it does not necessarily correspond to both $T^{\rm max}$ and
$\muB^{\rm max}$; this can be seen from the 2 GeV curve in
Fig.~\ref{fig_4}. We also see that the outgoing parts of the
trajectories at a given collision energy but different $\tauf$
sometimes do not overlap well, consistent with the significant
dependence of $\epsilon(t)$ and $\nB(t)$ on $\tauf$  at early times as
shown in Fig.~\ref{fig_1}. On the other hand, the returning parts of
the trajectories overlap well, because $\epsilon(t)$ and net
conserved-charge densities at late times are insensitive to $\tauf$. 

We have also calculated how the endpoint of a trajectory depends on
the collision energy at given formation times in Fig.~\ref{fig_4}.
At high collision energies, we observe a clear separation of the
endpoint curves for different $\tauf$ values. At very low collision
energies, however, this separation
decreases~\cite{Lin2018,Mendenhall2021}.
The sensitivity of the trajectory endpoint to $\tauf$ can be
understood in terms of the simpler uniform time
profile~\cite{Lin2018}, which have maximum values proportional  to
$\ln{(1+t_{21}/\tauf)}$. Note that the endpoint of a trajectory
corresponds to the maximum values of the energy and net-baryon
densities. At high collision energies where 
$t_{21}/\tauf \ll 1$, the maximum densities
are inversely proportional to $\tauf$ since
$\ln(1+t_{21}/\tauf) \sim t_{21}/\tauf$, so they are very sensitive to
the $\tauf$ value. 
On the other hand, at low collision energies where $t_{21}/\tauf \gg
1$, the maximum densities scale roughly as $\ln(t_{21}/\tauf)$, so
they are much less sensitive to the value of $\tauf$. Therefore, the
endpoint is more sensitive to the value of the formation time at high
collision energies than at low collision energies. 

At the threshold energy $E_0$, the endpoint curves at different
$\tauf$ converge to the same endpoint, 
which is located at $(\muB,T)\sim(900\,{\rm MeV},\ 60\,{\rm MeV})$. 
We also see that the CEP from the FRG
calculation~\cite{Fu2020} at $(\muB,T)=(635\,{\rm MeV},\ 107\,{\rm
  MeV})$ is well within the endpoint curves (even for $\tauf=0.9$
fm/$c$), which means that this CEP location should be accessible with
central Au+Au collisions. 
On the other hand, the $T-\muB$ region below the $\snn=2$ GeV
trajectory in the QCD phase diagram is essentially inaccessible to  
central Au+Au collisions according to our model. 

Note that our semi-analytical model should break down at very low
energies because the initial matter after the primary collisions would
not be in parton degrees of freedom. In fact, our semi-analytical
model predicts that the trajectories for ideal gas equations of state
start at the origin of the QCD phase diagram and return toward it
precisely because we assume the system is always in the parton
phase. It would be nice to further improve our model to
incorporate the energy loss of the participant nucleons 
or the interaction among secondary particles, which are beyond the
scope of this study. We have neglected secondary particles and their
interactions in order to analytically solve the resulting equations 
and get a reasonable analytical solution of the trajectory. 
On the other hand, these effects have been included in the AMPT model
calculations of the collision trajectories~\cite{Wang:2021owa}. 
Those AMPT model results and our semi-analytical results share many of
the same qualitative features of the trajectories for ideal gas
EoS.

Since our model gives $\epsilon^{\rm max}=2\rho_0\mN$ and
$\nB^{\rm max}=2\rho_0$ at the threshold energy as one naively
expects, the $\epsilon$ and $\nB$ values at very low energies from our
model should not be far off.
However, below the crossover curve, one expects the system to freeze out
and behave like a hadron resonance gas.
Therefore, the trajectories from the EoS of an ideal gas of massless
partons well below the crossover curve (or the first-order phase
transition line beyond the CEP) are not reliable.

\subsection{Trajectory of a massless QGP with the Boltzmann EoS}
\label{results_3}

\begin{figure*}
\includegraphics[width=\linewidth]{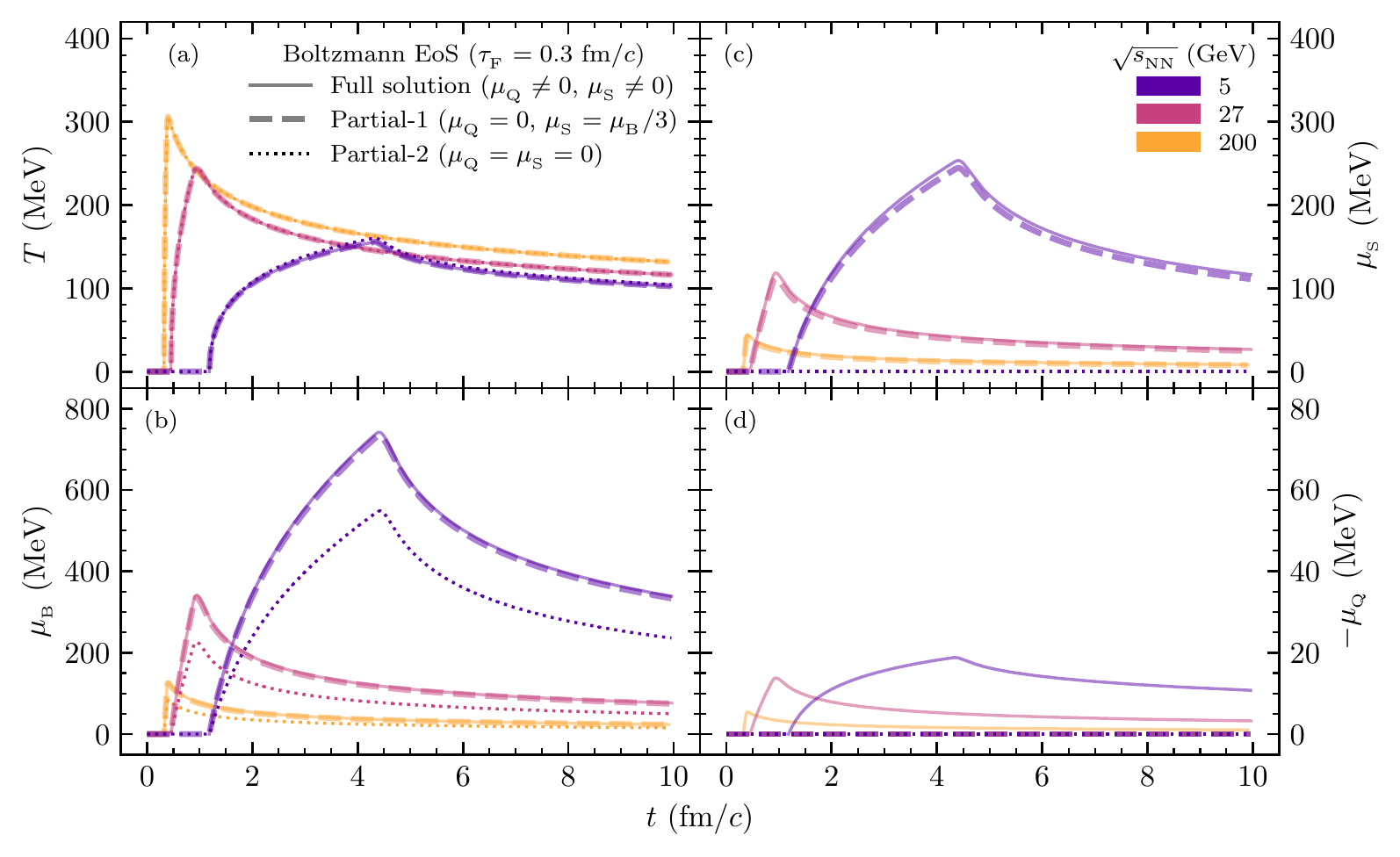}
\caption{Same as Fig.~\ref{fig_2}, but for the Boltzmann EoS and
  without the lattice EoS results.}
\label{fig_5}
\end{figure*}

We show in Fig.~\ref{fig_5} the
time evolutions of $T$ and $\mu$ extracted using the full solution,
the partial-1 solution, and the partial-2 solution 
of the Boltzmann ideal gas EoS for a massless QGP.
Similar to the results for the quantum EoS shown in
Fig.~\ref{fig_2}, the full and partial-1 solutions give essentially
the same results, while the partial-2 solution gives significantly
smaller $\muB$ values. In addition, the
magnitudes of $\muQ$ from the full solution of the Boltzmann EoS
are even smaller than those for the quantum EoS. As a result,
$\muS\simeq\muB/3$ also holds for the full solution here.

\begin{figure}
\includegraphics[width=\linewidth]{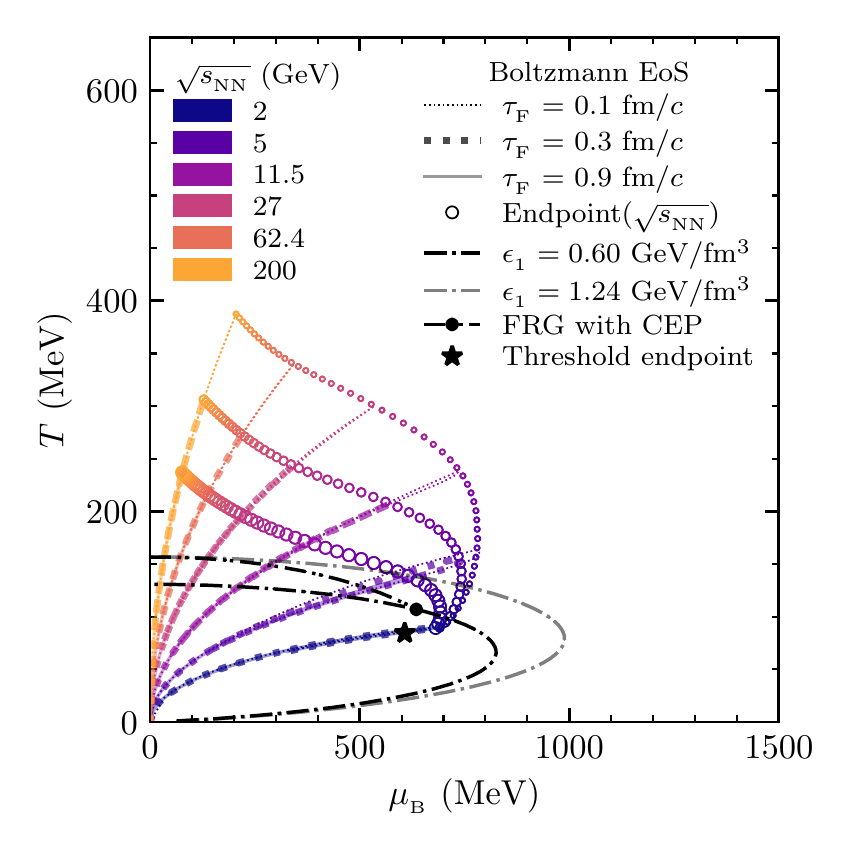}
\caption{Same as Fig.~\ref{fig_4}, but for the Boltzmann EoS.}
\label{fig_6}
\end{figure}

Figure~\ref{fig_6} shows how the $T-\muB$ trajectories using the full
solution of the Boltzmann EoS depend on $\tauf$. We observe the same
behavior as for the quantum EoS. 
The trajectory endpoint at the threshold energy is obtained by
using the partial-1 solution for the Boltzmann EoS with
$\epsilon^{\rm max}=2\rho_0\mN$ and $\nB^{\rm max}=2\rho_0$,
and it now lies to the left of the CEP.
We also show the two lines of constant $\epsilon$ as calculated from
the partial-1 solution of Eq.\eqref{partial_1_b_e} for the Boltzmann
EoS, which go through the endpoints of the FRG crossover curve.
Interestingly, they both show a half-loop structure, which is totally
different from the shape of constant-$\epsilon$ lines for the quantum
EoS. We can understand this by considering the total differential for
Eq.\eqref{partial_1_b_e}:
$d\epsilon_{_1}=\partial_{T}\epsilon_{_1}dT
+\partial_{\muB}\epsilon_{_1}d\muB$.
At constant energy density where $d\epsilon_{_1}=0$, we find
\begin{equation}
\frac{d\muB}{dT}=-\frac{\partial_T\epsilon_{_1}}{\partial_{\muB}\epsilon_{_1}}.
\end{equation}
The numerator in the above equation is zero when $\muB/(3T)\approx4.15$,
which corresponds to the turning point in the line of constant 
energy density for the Boltzmann EoS.
If one follows a line of constant $\epsilon$ starting from the
higher-temperature side at $\muB=0$ (where $\epsilon\gg\nB\mN$),
one will arrive at a point (near the turning point) beyond which
$\epsilon<\nB\mN$.
Since it is unnatural for a nuclear matter including the QGP to have
$\epsilon<\nB\mN$ if baryon number is effectively related to the
baryon mass, the $T-\muB$ points on the lower part (roughly the lower
half) of each constant-$\epsilon$ curve in Fig.~\ref{fig_6} do not
represent a physical QGP system with the Boltzmann EoS.

Figure~\ref{fig_6} also shows how the endpoint of a trajectory
depends on the collision energy at a given formation time.
As in Fig.~\ref{fig_4}, at high collision energies we see a clear
separation of the endpoint curves for different formation times.
Notably, the CEP from the FRG calculation is now much closer to the
$\tauf=0.9$ fm/$c$ endpoint curve than the  quantum EoS case. Another
key difference between Figs.~\ref{fig_6} and \ref{fig_4} is the
location of the CEP with respect to individual trajectories. 
For the Boltzmann EoS shown in Fig.~\ref{fig_6}, the CEP lies close to
the returning part of the 3.2 GeV trajectory or the outgoing part of
the 2.8 GeV trajectory. For the quantum EoS shown in Fig.~\ref{fig_4},
the CEP is close to the returning part of the 4.4 GeV trajectory or
the outgoing part of the 3.6 GeV trajectory.

\begin{figure}
\centering
\includegraphics[width=\linewidth]{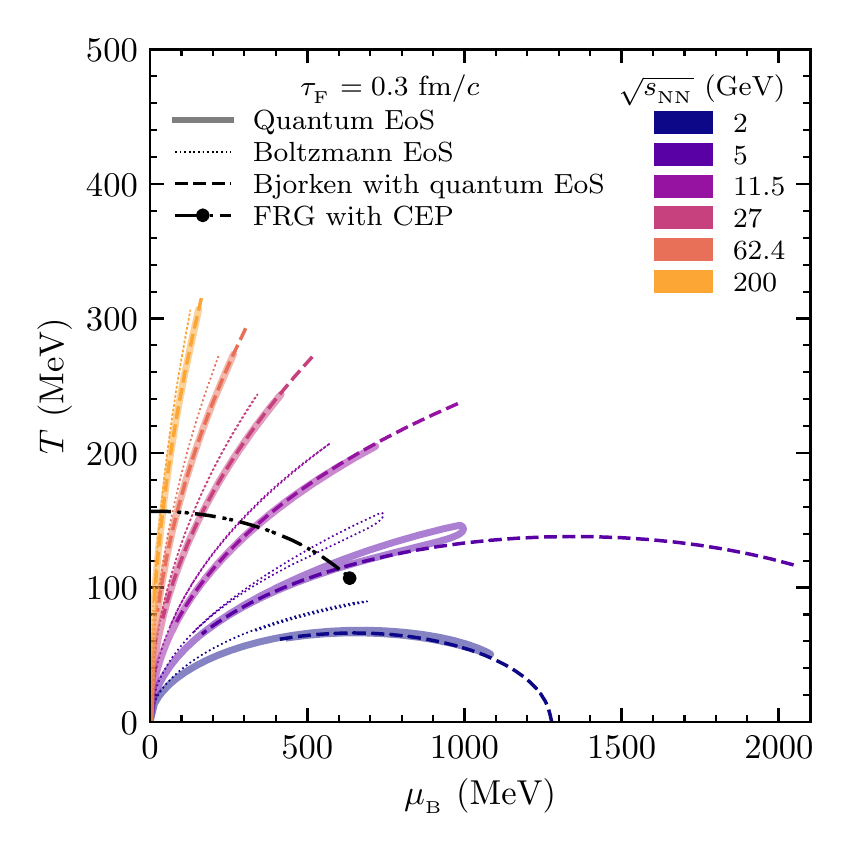}
\caption{Trajectories for a massless QGP from the quantum EoS (solid)
  and the Boltzmann EoS (dotted) using our $\epsilon(t)$ and $\nB(t)$
  in comparison with those from the quantum EoS using 
  $\epsilon^{_{Bj}}(t)$ and $\nB^{_{Bj}}(t)$ (dashed) at $\tauf=$ 0.3
  fm/$c$ for central Au+Au collisions at various energies. The 
  FRG crossover curve (dash-dot-dot) with the CEP is also shown
  for reference.}
\label{fig_7}
\end{figure}

In Fig.~\ref{fig_7}, we compare our results with the $T-\muB$
trajectories that are extracted from the $\epsilon(t)$ and $\nB(t)$
values calculated with the Bjorken formula~\cite{Bjorken1983}.
At high collision energies, the trajectories from the Bjorken formula
are rather close to our results that include the finite nuclear
thickness for the same (quantum) EoS, as expected.
At lower collision energies, however, $\muB^{\rm max}$ from the
Bjorken formula is much larger, and there is no outgoing part of the 
trajectory since $\epsilon^{\rm max}$ and $\nB^{\rm max}$ in the Bjorken
formula occur at the earliest time ($t=\tauf$). In contrast, in our
semi-analytical model the time of maximum density occurs much 
later, sometime within the range $[t_a,t_2+\tauf)$~\cite{Mendenhall2021}. 
We also observe in Fig.~\ref{fig_7} that the late-time part of a
trajectory from the Bjorken formula overlaps with the returning part of our
trajectory. This is because the Bjorken formula and our
semi-analytical model predict the same densities at late
times~\cite{Lin2018,Mendenhall2021}, which can be seen in
Fig.~\ref{fig_1}.

At low collision energies we observe a problem in extracting
the trajectory when using the Bjorken densities and the quantum EoS,
in that sometimes no solution for the $T-\muB$ trajectory exists at early
times. For example, at $\snn=2$ GeV and $\tauf=0.3$ fm/$c$, no
solution is found before $t\sim 8$ fm/$c$, at which time $\muB^{\rm
  max}\sim 1300$ MeV and  $T\sim 0$ MeV; afterwards the Bjorken
trajectory rises in $T$ and decreases in $\muB$ approaching the
returning part of the trajectory that is extracted using the densities
from our semi-analytical model. 
This problem occurs more often at low energies and small formation 
times, while it does not occur for collision energies higher than $\sim
5$ GeV at $\tauf=0.3$ fm/c. Note that this problem also does not occur
for the Boltzmann EoS~\cite{Lin:2022pem}. 
To understand this problem, we can use the partial-1 solution of the
quantum EoS in Eqs.\eqref{partial_1_q_e}-\eqref{partial_1_q_nB}, 
which give $\epsilon^{1/4}/n_{\rm _B}^{1/3} \geq
(2187\pi^2/128)^{1/12} \simeq 1.533$, 
where the equal sign corresponds to the solution at $T=0$. 
On the other hand, since the peak Bjorken energy and net-baryon
densities both change as $1/\tauf$, $\epsilon^{1/4}/n_{\rm _B}^{1/3}
\propto \tauf^{1/12}$ will decrease to the above value of 1.533 at a
certain finite $\tauf$ value. When we decrease $\tauf$ further,
there will be no $(T \geq 0,\muB)$ solution anymore.
Note that these scaling relations of the Bjorken density formulae are
not all the same as the those in the Bjorken hydrodynamics picture
that has $n\propto 1/\tauf$ but $\epsilon\propto 1/\tauf^{4/3}$. 

We also see in Fig.~\ref{fig_7} that the magnitude of the net-baryon
chemical potential can be large, even higher than 2 GeV. 
Using the Bjorken densities and the partial-1 solution of the quantum
EoS in Eqs.~(19)-(20),  one can show analytically that the baryon
chemical potential has no upper limit in the parton phase. As an
example, let us consider the peak Bjorken densities at $\tauf=0.3$
fm/$c$ at 27 GeV, which essentially corresponds to the endpoint of the
dashed magenta trajectory in Fig.~\ref{fig_7}.  
If we decrease $\tauf$, it is straightforward to show with
Eqs.~(19)-(20) that the trajectory endpoint will move to
the right to reach a higher $\muB$ while the $T/\muB$ ratio will
decrease. Therefore, the maximum  baryon chemical potential $\muB^{\rm
  max}$ for a Bjorken trajectory at a given energy (for all $\tauf$ 
values) is reached at $T=0$, where the partial-1 solution of the
quantum EoS in Eqs.~(19)-(20) gives 
\begin{equation}
\frac{\epsilon}{\nB}=\frac{3}{4} \muB^{\rm max}.
\end{equation}
In our model, the Bjorken energy and net-baryon
densities satisfy the following inequality because of Eq.(4):
\begin{equation}
\frac{\epsilon}{\nB}=\mN+\frac{dE_{\rm _T}/dy}{dN_{\rm netB}/dy} \geq \mN. \nonumber
\end{equation}
The above two relations then give
\begin{equation}
\muB^{\rm max}\geq 4\mN/3 \simeq 1.25 {~\rm GeV}. \nonumber
\end{equation}
Indeed, we can see from Fig.~\ref{fig_7} that the 2~GeV Bjorken trajectory
reaches the $T=0$ axis at a $\muB^{\rm max}$ value close to 1.25 GeV. At
higher energies, the $\muB^{\rm max}$ value (at $T=0$) will be even 
higher.

Comparing our results for the two ideal gas equations of state in
Fig.~\ref{fig_7}, we see that, while the $T^{\rm max}$ values are
often similar at the same collision energy (except at very low
energies), the $\muB^{\rm max}$ value is significantly larger in the
quantum EoS than in the Boltzmann EoS. This feature is also seen in
numerical results from the AMPT model~\cite{Wang:2021owa} and can be
understood in terms of the Pauli exclusion principle in quantum
statistics. In terms of the thermodynamics relations, this can be
understood by examining the first two terms in the Taylor expansion of
Eq.\eqref{partial_1_b_nB}: $n_{\rm _{B,1}}\simeq8\muB
T^2/(3\pi^2)+4\muB^3/(81\pi^2)$. The coefficient of each term in the
above equation is larger than the corresponding coefficient in
Eq.\eqref{partial_1_q_nB} for the quantum EoS. As a result, when the
same $\nB$ is used and the $T$ values are similar for the two
equations of state, $\muB$ for the quantum EoS will be larger.

\subsection{Trajectories of the lattice-QCD EoS}
\label{results_4}

In Fig.~\ref{fig_2}, we see that the lattice EoS results have a larger
$T$ than the results from the ideal gas EoS, which will lead to significantly
longer QGP lifetimes as we shall show in Fig.~\ref{fig_10}. At low collision
energies, no $T-\mu$ solution for the lattice EoS can be found
that corresponds to the $\epsilon$ and $\nB$ values at certain times.
This can be clearly seen in the 5 GeV results, which do not have
solutions at early times or late times. Moreover, the 5 GeV solutions
that are found  have a $\muB/T>2.5$ (as indicated by circles on the
curve), where we expect the lattice EoS to be
unreliable~\cite{Noronha-Hostler:2019ayj}.  Note that $\muB$ can first 
decrease at very early times because the lattice EoS trajectories
typically start at a finite $(\muB,T)$ instead of the origin. We also
see in Fig.~\ref{fig_2} that, while the $\muS$ value extracted with
the lattice EoS is reasonably close to the full solution of the
quantum EoS, the value of $\muQ$ from the lattice EoS is significantly
larger.

\begin{figure}
\includegraphics[width=\linewidth]{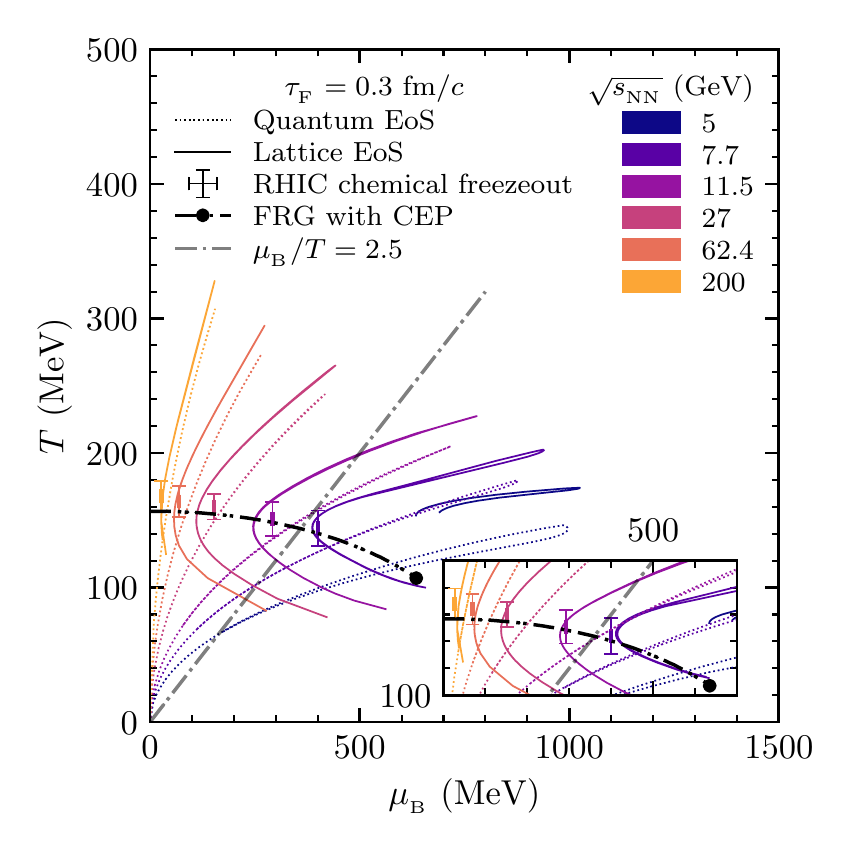}
\caption{Trajectories from the quantum EoS (dotted)
  and the lattice EoS (solid) at $\tauf=$ 0.3 fm/$c$
for central Au+Au collisions at various energies. 
The FRG crossover curve (dash-dot-dot) with the CEP (circle) and the
$\muB/T=2.5$ line are shown for reference, and  
the error-bar markers represent the RHIC chemical freezeout data. 
The inset shows the trajectories using the previous
$\dNBdytext$ parameterization~\cite{Mendenhall2021}.}
\label{fig_8}
\end{figure}

We compare in Fig.~\ref{fig_8} the trajectories extracted using the
full solutions of the quantum EoS and the lattice QCD-based EoS for 
$\tauf=0.3$ fm/$c$ for central Au+Au collisions. 
Compared to the quantum EoS, we see that $T^{\rm max}$ from the
lattice EoS is larger for all collision energies,
while $\muB^{\rm max}$ is slightly smaller at high collision energies
but significantly larger at low collision energies. 
In addition, the point at which the $T-\muB$ trajectory intersects the
FRG crossover curve shifts to smaller $\muB$, where the shift at high
energies is less noticeable than that at lower energies. 
Furthermore, the trajectory from the lattice EoS bends below the
crossover curve because of the smooth transition of the lattice EoS 
to the hadron resonance gas model at low temperatures.

Figure~\ref{fig_8} also shows the chemical freezeout
data~\cite{STAR:2017sal} extracted from grand canonical fits to the
particle yields, and we see that they are rather close to the
intersection points of our lattice EoS trajectories with the FRG
crossover curve. 
We note that the trajectories and intersection points depend on 
the parameterization of $\dNBdytext$. 
In the inset of Fig.~\ref{fig_8}, we show the trajectories that used
the previous $\dNBdytext$ parameterization in
Ref.~\cite{Mendenhall2021}, which appear to be closer to the RHIC
chemical freezeout data. 
In the region of small $\muB/T$, which corresponds to moderate to high
energies, the lattice EoS trajectories should be more realistic than 
trajectories from the quantum or Boltzmann ideal gas EoS. 
In the region of large $\muB/T$, however, the lattice EoS is expected
to be unreliable, for example, we see in Fig.~\ref{fig_8} that the
trajectory from the lattice EoS at $\snn=5.0$ GeV has no part below
the FRG crossover curve. In such cases, which corresponds to low
energies, the ideal gas EoS can still be used to calculate $T-\muB$ 
trajectories.

\subsection{Transverse Expansion}
\label{results_5}

To investigate the effect of transverse expansion, we first assume
that the radius of the transverse area $R_{\rm _T}$ in central A+A
collisions increases with time as
\begin{equation}
R_{\rm _T}(t)=R_A+\beta_{\rm _T}(t)\left(t-t_1-\tauf\right), 
\label{rT}
\end{equation}
where the transverse flow velocity is modeled as 
\begin{equation}
\beta_{\rm _T}(t)=
\begin{cases}
0,& \text{ for }t<t_1+\tauf \\
\left[1-e^{-(t-t_1-\tauf)/t_{\rm _T}}\right]\beta_{\rm _{T,f}},&
\text{ for }t \geq t_1+\tauf.
\end{cases}
\label{beta_T}
\end{equation}
We parameterize the final value of the transverse flow velocity
$\beta_{\rm _{T,f}}$ as 
\begin{equation}
\beta_{\rm _{T,f}}=\left[\frac{\ln\left(\snn/E_0\right)}{64.7 + \ln\left(\snn/E_0\right)}\right]^{0.202}.
\end{equation}
The above parameterization uses the kinetic freezeout parameters
obtained by fitting the transverse momentum spectral shapes 
in central Au+Au collisions from 7.7 to 200 GeV 
and Pb+Pb collisions at 2.76 TeV to a blast-wave
model~\cite{STAR:2008med,STAR:2017sal,ALICE:2013mez}. 
For the low energy behavior, we assume
$\beta_{\rm _{T,f}}\to0$ as $\snn\to E_0$ and use the kinetic
freezeout parameters reported in Ref.~\cite{STAR:2017sal}, which
contained various
data~\cite{E802:1996owm,E-802:1998xum,E866:1999ktz,E877:1999qdc,E802:1999hit,E-802:1999ewk,E866:2000dog,Klay2002,NA49:2007stj,NA49:2006gaj,NA49:2004mrq,NA49:2002pzu}. 

Next, we parameterize $t_{\rm _T}$, the timescale for the development
of transverse flow, by assuming $t_{\rm _T}\propto 1/n^{\rm  max}$,
with $n^{\rm  max}$ being the parton number density at $\epsilon^{\rm
  max}$, since the mean-free-path of a parton is inversely
proportional to the parton number density. 
For simplicity, we calculate $n^{\rm max}$ with the Boltzmann EoS relation, 
$n=\sqrt[4]{52\epsilon^3/27/\pi^2}$, and adopt the following
$\epsilon^{\rm max}$ that assumes a uniform production profile in
time~\cite{Lin2018}:
\begin{equation}
\epsilon^{\rm max}_{uni}=\frac{1}{\pi R_A^2 (t_2-t_1)} \frac{dm_{\rm _T}}{dy}(0)
~\ln \! \left(1+\frac{t_2-t_1}{\tauf} \right).
\end{equation}
Note that in the above equation we take $t_1$ and $t_2$ as
0.264$d_t$ and $0.736d_t$, respectively, so that the uniform time
profile matches the mean and standard deviation of time as the 
uniform $g(z_0,x)$ and Eq.\eqref{t1t2} used by our semi-analytical
model. We then normalize $t_{\rm _T}$ to a given value $t_{\rm _T}^{\rm
  norm}$ at $\snn=200$ GeV and $\tauf=0.3$ fm/$c$, and we shall vary
$t_{\rm _T}^{\rm norm}$ from 1 to 6 fm$/c$.

\begin{figure}
\centering
\includegraphics[width=\linewidth]{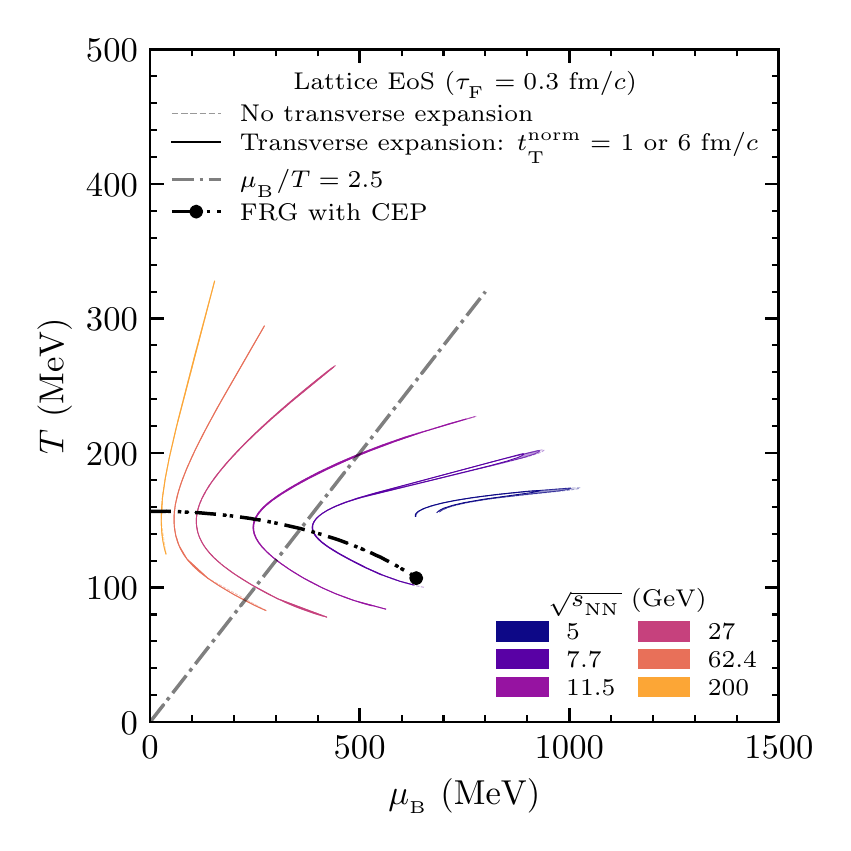}
\caption{Trajectories from the lattice EoS without (dashed) or with
  transverse expansion (solid) at $\tauf=$ 0.3 fm/$c$ for Au+Au
  collisions at various energies. The FRG crossover curve
  (dash-dot-dot) with the CEP (circle) and the $\muB/T=2.5$ line are
  also shown for reference.}
\label{fig_9}
\end{figure}

Figure~\ref{fig_9} shows the $T-\muB$ trajectories after including the
transverse expansion with $t_{\rm _T}^{\rm norm}$ of 1 or 6 fm$/c$
(solid), where the area between the two solid curves at each
collision energy is shaded to show the range of the effect of
transverse expansion. The trajectories without transverse expansion
are also shown (dashed), which corresponds to the case of transverse
expansion with $t_{\rm _T}^{\rm norm} \rightarrow \infty$. 
We see that the path of the trajectory in the QCD phase
diagram changes little when including the effect of 
transverse expansion for all collision energies. 
Instead, the endpoint of the trajectory shifts closer to the origin
(more noticeable at low energies), qualitatively similar to the effect
of a larger $\tauf$.

The small effect of transverse expansion on the trajectories can be
understood as follows. Our implementation of transverse expansion 
is done via  a time-dependent transverse area $A_{\rm _T}(t)$ that increases
with a timescale $t_{\rm _T}$. 
At early times (i.e., when $t-t_1-\tauf < t_{\rm _T}$), the transverse flow
has not developed much, thus transverse expansion has little effect on
the densities and the trajectory (typically the outgoing part). 
At late times (i.e., when $t-t_1-\tauf>t_{\rm _T}$), transverse expansion 
decreases the energy density and net-baryon density by the same
factor. When we neglect the  transverse expansion, the late
time evolution of densities in our model approaches that from the
Bjorken formula and decrease as $1/t$; therefore they also decrease
with time at the same rate. 
This means that the late time densities (and thus the trajectory
point) with transverse expansion at time $t$ is the same as the
trajectory without transverse expansion at a later time
$t'$ (with $t'>t$). Therefore, the returning part of  a given
trajectory with transverse expansion overlaps with the trajectory
without transverse expansion, and the main effect of transverse
expansion is on the turning point of the trajectory. 

\subsection{QGP lifetime}
\label{results_6}

\begin{figure*}
\includegraphics[width=\linewidth]{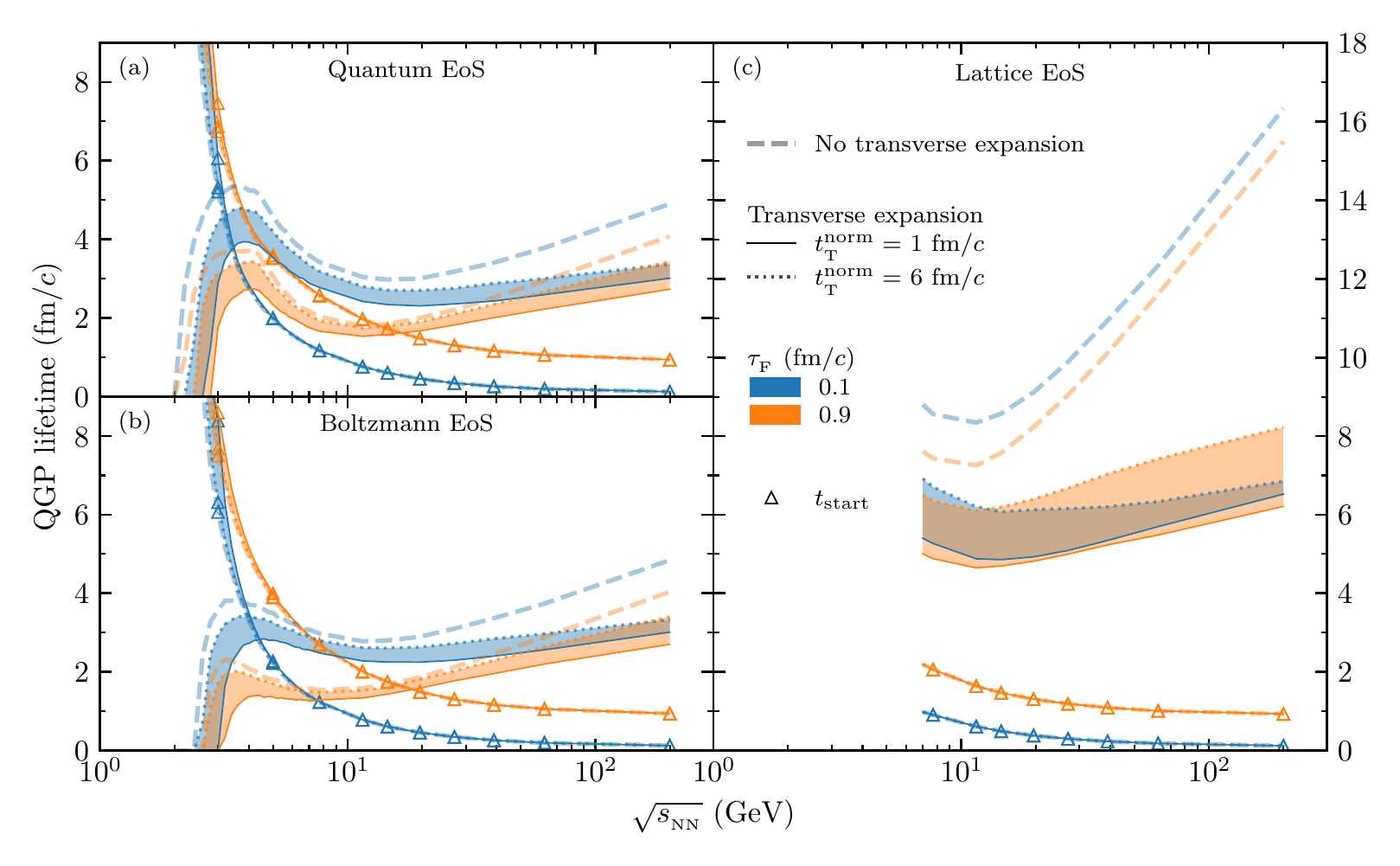}
\caption{The starting time (curves with triangles) and lifetime
  (curves without symbols) of the QGP phase in central Au+Au
  collisions from (a) the quantum EoS, (b) the Boltzmann EoS, and (c)
  the lattice EoS. Results are for $\tauf=$ 0.1 (blue) and
  0.9 fm/$c$ (orange) without transverse expansion (dashed) and
  with transverse expansion at $t_{\rm _T}^{\rm norm}=1$ (solid)
  and 6 fm$/c$ (dotted).}
\label{fig_10}
\end{figure*}

We can calculate the time when the matter enters the parton phase
($t_{\rm start}$), the time when it exits the parton phase ($t_{\rm end}$), and
the QGP lifetime as $\tQGP=t_{\rm end}-t_{\rm start}$. 
Specifically, we calculate the first and last
times when a $T-\muB$ trajectory intersects the FRG crossover
curve~\cite{Fu2020}. For very low collision energies, the trajectory
does not intersect the FRG crossover curve, so we calculate
$t_{\rm start}$ and $t_{\rm end}$ by finding when the trajectory intersects
the line of constant energy density $\epsilon=0.51$ GeV/fm$^3$ shown
in Fig.~\ref{fig_3}. From the results shown in Fig.~\ref{fig_10}, we
see that $t_{\rm start}$ at a given energy is larger for a larger
$\tauf$ as expected, and $t_{\rm start}$ is significantly larger at
lower energies mostly due to the longer crossing time $d_t$.  
From the results without transverse expansion (dashed curves), 
we see that the QGP lifetime is shorter for a larger $\tauf$, 
since $t_{\rm end}$ (and the late-time trajectory) is almost
independent of $\tauf$ without transverse expansion. 
In addition, the QGP lifetimes from the lattice EoS are much larger
than those from the quantum or Boltzmann EoS, mainly because the
late-time temperatures from the lattice EoS are significantly higher
as shown in Fig.~\ref{fig_2}.

When transverse expansion is considered, we
see in Fig.~\ref{fig_10} that the QGP lifetime becomes significantly
shorter, especially at high energies. Note that the area between the
$t_{\rm _T}^{\rm norm}=1$ and 6 fm$/c$ curves at each given $\tauf$ is
shaded to show the range of transverse  expansion effect, where a
larger  $t_{\rm _T}^{\rm norm}$ value makes the transverse flow
development slower and consequently increases the QGP lifetime.  
The $t_{\rm start}$ values (curves with triangles) are essentially
unaffected by the transverse expansion, because the transverse
expansion takes some time to develop. However, the transverse
expansion leads to lower densities at late times and thus 
decreases $t_{\rm end}$ and the QGP lifetime.  For example, at
$\snn=200$ GeV the QGP lifetime from the lattice EoS is $\tQGP \sim 7$
fm/c with transverse expansion but $\sim 16$ fm/c without. 
Therefore, the transverse expansion has a large effect on the
QGP lifetime at all energies, although it has a little effect on the
collision trajectory including its distance to the CEP at energies
above $\sim 7.7$ GeV. We also see that $\tQGP$ is not always larger
for a smaller $\tauf$ (more obvious for the lattice EoS
results), unlike the case without transverse expansion. This behavior
may seem counterintuitive at first, but it occurs because a smaller
$\tauf$ leads to higher densities and consequently a faster transverse
expansion, which leads to a smaller $t_{\rm end}$. 

We see in Fig.~\ref{fig_10} that the QGP lifetime does not increase
monotonously with the collision energy, as one naively expects. 
Instead, $\tQGP$ may have a local maximum at low collision energies,
which is between 3 and 5 GeV from the ideal gas EoS results shown in
Figs.~\ref{fig_10}(a)-(b). 
Unfortunately, the lattice EoS results cannot reach very low energies
due to the high $\muB/T$ value there, although they also hint at an 
increase of $\tQGP$ as $\snn$ decreases below 11.5 GeV.
Note that a numerical study with the AMPT model~\cite{Wang:2021owa}
has also observed the non-monotonous dependence of $\tQGP$ on 
the collision energy for the Boltzmann EoS. 
After reaching the local maximum, the QGP lifetime decreases with the
collision energy before it increases again, where the increase
at high collision energies (after considering the transverse expansion)
is rather slow for the lattice EoS. 
We also note that the STAR Collaboration recently showed that the
matter produced in Au+Au collisions at $\snn=3$ GeV is likely
dominated by baryonic interactions~\cite{STAR:2021yiu}. 
Our results in Figs.~\ref{fig_10}(a)-(b) indicate that 
whether the QGP could be produced in such collisions
depends on the details such as the formation time, the equation of
state, and the rate of transverse expansion. At $\snn=4$ GeV, on the
other hand, the QGP will be formed independent of these details
according to our results.

\section{Discussions}
\label{discussions}

\begin{figure*}
\centering
\includegraphics[width=\linewidth]{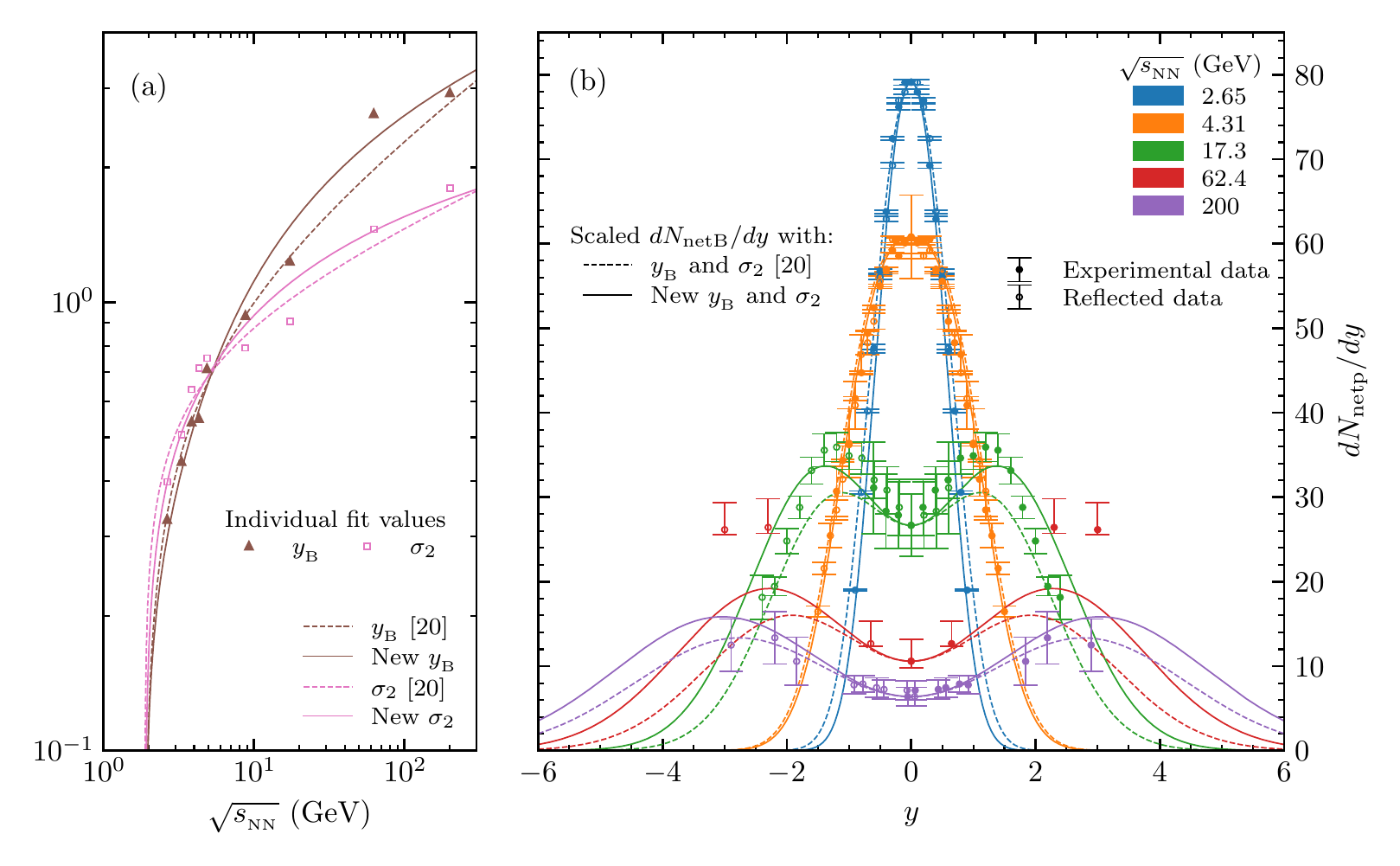}
\caption{(a) Individual fit values of $y_{\rm _B}$
  (triangles) and $\sigma_2$ (squares) together with their previous 
  parameterizations~\cite{Mendenhall2021} (dashed) and 
  new parameterizations in Eq.~\eqref{ybs2} (solid). (b) the
  experimental $dN_{\rm netp}/dy$ data (filled circles) and the
  values reflected across mid-rapidity (open-circles) together with
  scaled $\dNBdytext$ at $\snn=$ 2.65, 4.31, 17.3, 62.4, and 200 GeV
  using either the parameterizations from Ref.~\cite{Mendenhall2021}
  (dashed) or the new parameterizations from Eq.\eqref{ybs2}
  (solid).} 
\label{fig_11}
\end{figure*}

\begin{figure}
\centering
\includegraphics[width=\linewidth]{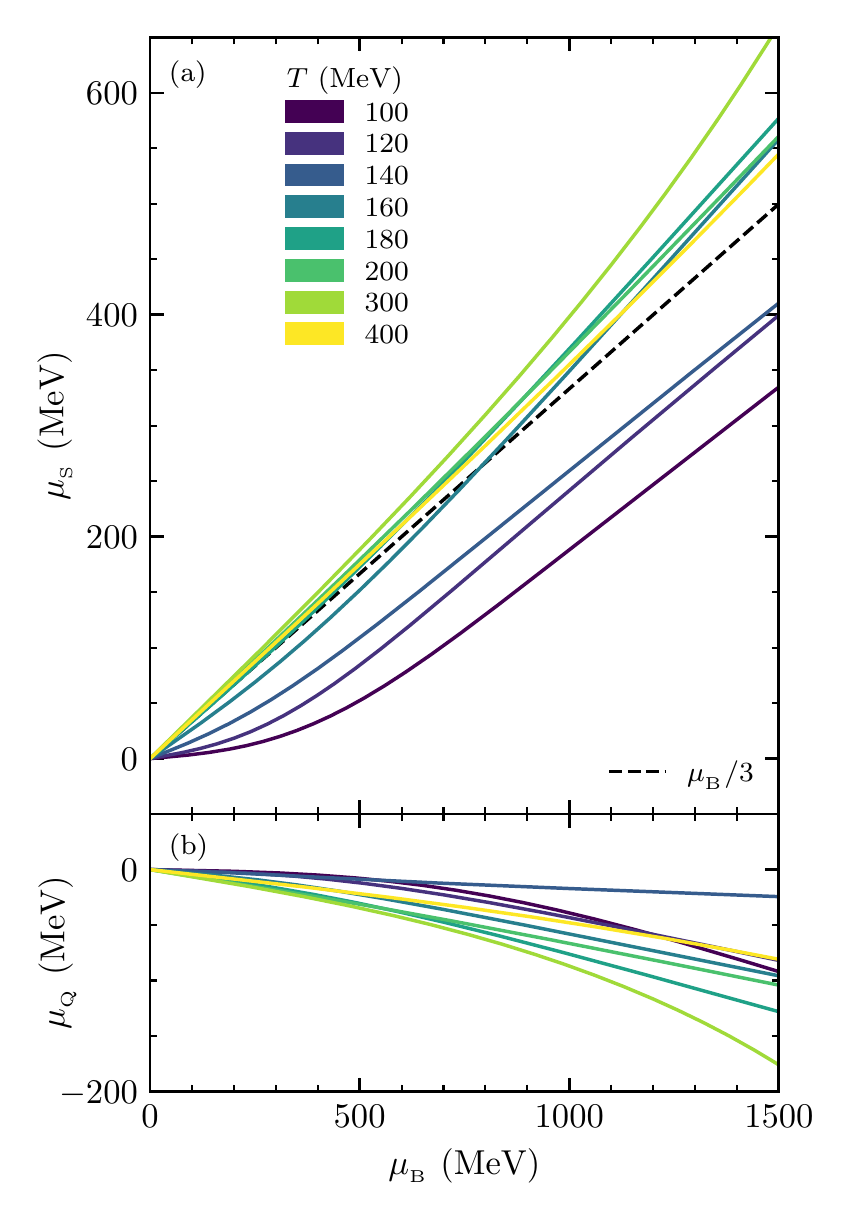}
\caption{Chemical potential (a) $\muS$ and (b) $\muQ$ versus $\muB$
  at various temperatures for the lattice EoS under the condition
  $\nQ=\nB Z/A$ (for Au) and $\nS=0$. The dashed line
  represents $\muS=\muB/3$.}
\label{fig_12}
\end{figure}

In this study we use new parameterizations for $y_{\rm _B}$ and
$\sigma_2$ to describe $\dNBdytext$. 
We have made this change for two reasons. First, the previous  
parameterizations~\cite{Mendenhall2021,Lin:2022pem} contain a positive
power of $\snn$ e.g., $y_{\rm _B} \propto (\snn-E_0)^{\!0.196}
\ln^{0.392}\snn$. This functional form will break down above a certain
high energy because $y_{\rm _B}>\yCM$ will eventually cause the
energy in the $\dNBdytext$ term of Eq.\eqref{dmtdy} to exceed the total
energy, which would be unphysical. We thus use the new forms in
Eq.~\eqref{ybs2} that do not contain positive powers of $\snn$. 
Second, we have corrected the collision energies for the low energy
proton $dN/dy$ data~\cite{Klay2002}, from $\snn=$ 2.4, 3.1, 3.6, and
4.1 GeV~\cite{Mendenhall2021} to $\snn=$ 2.65, 3.31, 3.85, and 4.31
GeV, after using the actual beam kinetic energies after correcting for
the energy loss before reaching the target~\cite{Klay2002}. 
We also realized that the net-proton $dN/dy$ data at $\snn~5$ GeV 
contain data at $\snn=$ 4.70~\cite{E877:1999qdc},
4.86~\cite{E802:1999hit}, and 4.88 GeV~\cite{E917:2000spt}. Therefore, 
we now combine the data at 4.86 GeV and 4.88 GeV into one data set at
$\snn=4.87$ GeV. In addition, we now include the net-proton data at
$\snn=$ 8.77~\cite{NA49:2010lhg} and 62.4 GeV~\cite{BRAHMS:2009wlg}. 

In Fig.~\ref{fig_11}(a), we show the updated individual fit values for
$y_{\rm _B}$ and $\sigma_2$ in comparison with the
old~\cite{Mendenhall2021} and new parameterizations. 
At $\snn<4$ GeV, the new parameterizations  
for $y_{\rm _B}$ and  $\sigma_2$ match the individual fit values quite
well. On the other hand, the individual fit values at $\snn>4$
GeV for $y_{\rm _B}$ or $\sigma_2$ are such that it seems impossible to
fit them well with a smooth function; Instead, our parameterization
provides a smooth fit that is overall relatively close to most
individual fit values. 
For example, the new parameterization overestimates the individual
$y_{\rm _B}$ values at $\snn=$ 8.8 and 17.3 GeV, which leads to an
underestimate of $\dNBdytext(0)$; this is the reason why the
trajectories around these energies intersect the FRG crossover curve 
at lower $\muB$ than the RHIC chemical freezeout data as shown in
Fig.~\ref{fig_8}. In Fig.~\ref{fig_11}(b), we compare the $\dNBdytext$
shape from both old and new parameterizations with the net-proton data 
at several energies. We note that the $\dNBdytext$ shape is more
sensitive to the $y_{\rm _B}$ parameter than $\sigma_2$.  
We see that both parameterizations fit well the
shape of the data at 2.65 and 4.31 GeV. At $\snn=17.3$ GeV, however,
the new parameterization peaks at a higher rapidity than the data,
which is consistent with its overestimate of the individual $y_{\rm
  _B}$ fit value as shown in Fig.~\ref{fig_11}(a). At 62.4 GeV the new
parameterization has a relatively lower peak $dN/dy$ value than the
data, consistent with its underestimate of the corresponding individual
$y_{\rm _B}$ fit value.

For the ideal gas equations of state, so far we have only considered a
QGP consisting of massless quarks and gluons. 
A more realistic approach would be to consider a massive 
$s$-quark, since $m_s=95$ MeV/$c^2$~\cite{Tanabashi2018} is not
negligible compared to the system's temperature scale.
We have not found an analytical solution for the
total energy density of massive $s$ and $\bar{s}$
with the quantum EoS; therefore, we create an interpolating
function of temperature:
\begin{equation}
	\epsilon_s+\epsilon_{\bar{s}}=\frac{6}{\pi^2}\int\limits_0^\infty
	\frac{p^2\sqrt{p^2+m_s^2}~dp}{e^{\sqrt{p^2+m_s^2}/T}+1}
	=\epsilon_{s+\bar{s}}(T).
	\label{s_Int}
\end{equation}
Note that we have used the relation $\mu_s=0$, which results from the
strangeness neutrality in our semi-analytical model. Considering the
finite $s$-quark mass, we obtain the following energy density:
\begin{equation}
  \begin{split}
    &\epsilon=\frac{37\pi^2}{30}T^4+3\frac{(\muB-2\muS)^2+\muS^2}{2}T^2 \\
    &\qquad+3\frac{(\muB-2\muS)^4+\muS^4}{4\pi^2}+\epsilon_{s+\bar{s}}(T).
  \end{split}
  \label{q_ms95_nQ}
\end{equation}
The equations for $\nB$ and $\nQ$ are unchanged from
Eqs.\eqref{full_q_nB} and \eqref{full_q_nQ}, i.e., only the $\epsilon$
equation changes when considering $m_s\neq0$.  
We numerically solve these equations using enough $T$ sampling points
for $\epsilon_{s+\bar{s}}(T)$ to ensure accuracy. We have compared the resulting
$T-\muB$ trajectories
to those for $m_s=0$ MeV/$c^2$ and have observed essentially no
difference in the $T-\muB$ results including the QGP lifetime when
considering a non-zero $s$-quark mass. Note that a similar lack of
effect from finite quark masses have also been seen in numerical
studies from the AMPT model~\cite{Wang:2021owa}. 

We have observed $\muQ\simeq 0$ in the full solutions for both
quantum and Boltzmann ideal gas equations of state, as shown in
Figs.~\ref{fig_2} and \ref{fig_5}. As a result, the partial-1 solution
assumes $\muQ=0$ and then gives almost the same $T-\muB$ trajectories
as the full solution. Therefore, a natural question is why $\muQ$ is
so small. To answer this question, we first observe that 
our semi-analytical model~\cite{Mendenhall2021} gives $\nS(t)=0$. In
addition, our model predicts $\nQ(t)=\nB(t)/2$ if $Z=A/2$ for the
colliding nuclei.
One can verify that, $\nS=0$ and $\nQ=\nB/2$ lead to $\muQ=0$ for both
statistics in the ideal gas EoS. Therefore, $\muQ\simeq 0$ is a
consequence of the fact that most nuclei have $Z \sim A/2$.
Note that recent results from the AMPT model~\cite{Lin2005} also show
$\muQ\approx 0$~\cite{Wang:2021owa}, although the AMPT model does not
assume the $s$-$\bar s$ symmetry for the initial production and thus
does not give $\nS(t)=0$ exactly. 
We also note that the $\muQ$ values extracted with the lattice EoS are
sometimes significantly larger than those from the full solution of
the ideal gas EoS with either statistics. 

We examine in Fig.~\ref{fig_12} the strangeness neutrality in the
lattice EoS by plotting the $\muB$ dependence of $\muS$ and $\muQ$ for
various temperatures. Note that the lattice EoS results are obtained
under the condition of Eq.\eqref{condition} using the $Z$ and $A$ value 
for the gold nucleus. 
We see $\muS \simeq \muB/3$ for temperatures higher than $T\gtrsim160$
MeV. For lower $T$, however, this approximation does not describe the
lattice results, which are expected to be  trustworthy at least at low
$\muB$~\cite{Noronha-Hostler:2019ayj}. 
These features are very similar to the FRG results~\cite{Fu:2018swz}. 
The results from the FRG method also show a strict ordering of
$\muS(\muB)$ with $T$, where the $\muS$ result at higher $T$ gets
closer to the $\muS=\muB/3$ line. 
The lattice EoS results in Fig.~\ref{fig_12} show a similar
ordering of $\muS$ below $T \sim 180$ MeV but no clear
ordering at higher temperatures in either $\muS$ or $\muQ$. 
In addition, in the region $\muB/T<2.5$ we observe that $\muQ$ is
small with a magnitude not larger than $\sim 60$ MeV.  
Overall, the partial-1 assumptions of $\muQ=0$ and $\muS=\muB/3$ work
less well for the lattice EoS than for the ideal gas EoS. 
We find that for the lattice EoS the trajectories calculated with these
assumptions can have $T$ and $\muB$ values different by up to $\sim
2\%$ and $\sim 16\%$, respectively;  
interestingly, the calculated QGP lifetime is not much different. 

Since our model is semi-analytical, it is a convenient tool for
calculating the trajectories of nuclear collisions in the QCD phase
diagram, either in the conventional $T-\muB$ plane or the more
general $T-\muB-\muQ-\muS$ four-dimensional space. We have 
extended the web interface~\cite{interface}  that performs the
semi-analytical calculation of $\epsilon(t)$ for central A+A
collisions depending on the user input for the colliding system,
energy, and the proper formation time $\tauf$. 
It~\cite{interface} now performs the semi-analytical calculations of
$\epsilon(t)$, $n(t)$, $T(t)$, and $\mu(t)$ for either the quantum or
Boltzmann ideal gas equations of state and also plots the 
$T-\muB$ trajectory. We plan to further extend the web interface to
include the lattice EoS and transverse expansion.

\section{Conclusions}
\label{conclusions}

We have extended a semi-analytical model, which considers
the finite nuclear thickness, to calculate the energy density $\epsilon(t)$,
net-baryon density $\nB(t)$, net-charge density $\nQ(t)$,
and net-strangeness density $\nS(t)$ at mid-spacetime-rapidity
averaged over the full transverse area of central Au+Au collisions.
We then extract the temperature $T(t)$, baryon chemical potential
$\muB(t)$, electric charge chemical potential $\muQ(t)$, and strangeness
chemical potential $\muS(t)$ of the parton system assuming the
formation of an equilibrated QGP. 
We use an ideal gas equation of state of either quantum or Boltzmann
statistics or a lattice QCD-based equation of state. 
We find that the trajectory in the $T-\muB$ plane significantly
depends on the EoS; for example, the critical end point
from the FRG method is located close to the $\snn \sim 4$ GeV
trajectory when using the quantum ideal gas EoS but the $\snn \sim 3$ GeV
trajectory when using the Boltzmann ideal gas EoS. 

By calculating the trajectory endpoint as a function of collision
energy, we obtain the $T-\muB$ area that the mid-pseudorapidity
region of central Au+Au collisions can cover.  
We find that the accessible area in the phase diagram depends strongly
on the parton formation time $\tauf$, once the collision energy is
higher than several GeV. On the other hand, the critical end point
from the FRG calculation is well within the accessible area, even for
a large formation time of $\tauf=0.9$ fm/$c$. We also observe that the
trajectory using the Bjorken energy density method is significantly
different at low collision energies, which further demonstrates the
importance of the finite nuclear thickness at low energies such as the
BES energies. In addition, we find that the transverse expansion has a
small effect on the collision trajectory but a large effect on the QGP
lifetime, while the QGP lifetime for the lattice EoS
is much larger than those for the ideal gas EoS. 
We further observe an unexpected increase in the QGP lifetime
as the collision energy decreases below $\snn \sim 11.5$ GeV. 
Overall, our semi-analytical model provides a useful tool for 
exploring the trajectories of nuclear collisions in the 
QCD phase diagram in the $T-\muB$ plane or in the more general 
$T-\muB-\muQ-\muS$ space.

\begin{acknowledgments}
This work has been supported by the National Science Foundation
under Grant No. 2012947. We thank Wei-jie Fu, Guo-Liang Ma, 
Han-Sheng Wang, Claudia Ratti, and Christina Markert for helpful
discussions. We also thank Wei-jie Fu for providing the crossover and
CEP results from FRG calculations.
\end{acknowledgments}

\appendix*

\section{Thermodynamics relations between $\epsilon$, $n$ and $T$, $\mu$ for a QGP with ideal gas equations of state} 
\label{appendix}

We consider a quark-gluon plasma composed of gluons
$g$ and three quark flavors, similar to a previous
study~\cite{Wang:2021owa}.  The total energy, net-baryon, net-electric
charge, and net-strangeness densities are then given
by
\begin{align}
  &\epsilon=\epsilon_g+\sum_{q}\left(\epsilon_q+\epsilon_{\bar{q}}\right),
  &\nB=\sum_{q}B_q\left(n_q-n_{\bar{q}}\right), \nonumber \\
  &\nQ=\sum_{q}Q_q\left(n_q-n_{\bar{q}}\right),
  &\nS=\sum_{q}S_q\left(n_q-n_{\bar{q}}\right),
  \label{n_S}
\end{align}
where $B_q$, $Q_q$, and $S_q$ are the quark baryon, electric charge, and
strangeness numbers, respectively, for quark flavor $q$ with $q=u$,
$d$, or $s$. For parton flavor $i$, the energy and number densities
are given by 
\begin{align}
  \begin{split}
    &\epsilon_i=\frac{1}{2\pi^2}\int\limits_0^\infty dp~p^2
    \sqrt{p^2+m_i^2}~f_i(p), \\
    &n_i=\frac{1}{2\pi^2}\int\limits_0^\infty dp~p^2 f_i(p),
\end{split}
\end{align}
respectively, where $m_i$ represents the parton mass, and $f_i(p)$ is given by
\begin{equation}
f_i(p)=d_i \left[\exp\left(\frac{\sqrt{p^2+m_i^2}-\mu_i}{T}\right)+K\right]^{-1}.
\label{f_i}
\end{equation}
In the above, the degeneracy factor $d_i$ is 16 for gluons and 6 for
quarks, $\mu_i$ is the parton chemical potential, $T$ is the
temperature, $K=1$ for Fermi-Dirac statistics, $-1$ for
Bose-Einstein statistics, and 0 for Boltzmann statistics.  
The chemical potential of
parton flavor $i$ is $\mu_i=B_i\muB+Q_i\muQ+S_i\muS$,
where $\muB$, $\muQ$, and $\muS$ are the baryon, electric charge,
and strangeness chemical potentials, respectively.

\subsection{Massless QGP with quantum statistics}

For a massless quark-gluon plasma with quantum statistics,
we take $\mu_g=0$, $\mu_q+\mu_{\bar{q}}=0$ and then obtain
\begin{equation}
  \begin{split}
    \epsilon_q+\epsilon_{\bar{q}}&=-\frac{18T^4}{\pi^2}
    \left[Li_4\left(-e^{-\mu_q/T}\right)+Li_4\left(-e^{\mu_q/T}\right)\right] \\
    &=\frac{7\pi^2}{20}T^4+\frac{3\mu_q^2}{2}T^2+\frac{3\mu_q^4}{4\pi^2}, \\
    n_q-n_{\bar{q}}&=\frac{6T^3}{\pi^2}
    \left[Li_3\left(-e^{-\mu_q/T}\right)-Li_3\left(-e^{\mu_q/T}\right)\right] \\
    &=\mu_qT^2+\frac{\mu_q^3}{\pi^2},
  \end{split}
\end{equation}
where $Li_{\rm n}(z)$ is the polylogarithm function of order
n. Therefore, we obtain the following:
\begin{widetext}
\begin{align}
	\begin{split}
	\epsilon&=\frac{19\pi^2}{12}T^4
        +\frac{(\muB+2\muQ)^2+(\muB-\muQ)^2+(\muB-\muQ-3\muS)^2}{6}T^2 \\
        &\quad\quad+\frac{(\muB+2\muQ)^4+(\muB-\muQ)^4+(\muB-\muQ-3\muS)^4}{108\pi^2},
	\end{split}
	\label{ep_q} \\
    	\nB&=\frac{\muB-\muS}{3}T^2
    	+\frac{(\muB+2\muQ)^3+(\muB-\muQ)^3+(\muB-\muQ-3\muS)^3}{81\pi^2},
	\label{nB_q} \\
    	\nQ&=\frac{2\muQ+\muS}{3}T^2
    	+\frac{2(\muB+2\muQ)^3-(\muB-\muQ)^3-(\muB-\muQ-3\muS)^3}{81\pi^2},
  	\label{nQ_q} \\
  	\nS&=-\frac{\muB-\muQ-3\muS}{3}T^2-\frac{(\muB-\muQ-3\muS)^3}{27\pi^2}.
  	\label{nS_q}
\end{align}
\end{widetext}

\subsection{Massless QGP with Boltzmann statistics}

Using the Maxwell-Boltzmann statistics to describe the thermodynamics
of the quark-gluon plasma, for massless partons we obtain the
following equations that are simpler than those for finite quark
masses~\cite{Wang:2021owa}: 
\begin{widetext}
\begin{align}
	\epsilon&=\frac{12}{\pi^2}T^4\Bigg[4+3\cosh{\left(\frac{\muB+2\muQ}{3T}\right)}
	+3\cosh{\left(\frac{\muB-\muQ}{3T}\right)}
	+3\cosh{\left(\frac{\muB-\muQ-3\muS}{3T}\right)}\Bigg],
	\label{ep_b} \\
	\nB&=\frac{4}{\pi^2}T^3\Bigg[\sinh{\left(\frac{\muB+2\muQ}{3T}\right)}
	+\sinh{\left(\frac{\muB-\muQ}{3T}\right)}
	+\sinh{\left(\frac{\muB-\muQ-3\muS}{3T}\right)}\Bigg],
	\label{nB_b} \\
	\nQ&=\frac{4}{\pi^2}T^3\Bigg[2\sinh{\left(\frac{\muB+2\muQ}{3T}\right)} 
	-\sinh{\left(\frac{\muB-\muQ}{3T}\right)}
	-\sinh{\left(\frac{\muB-\muQ-3\muS}{3T}\right)}\Bigg],
	\label{nQ_b} \\
	\nS&=-\frac{12}{\pi^2}T^3\sinh{\left(\frac{\muB-\muQ-3\muS}{3T}\right)}.
	\label{nS_b}
\end{align}
\end{widetext}

\bibliography{traj}

\end{document}